# Chemical Abundances of a Sample of Oxygen-dominated Galaxies

B. E. Miranda-Pérez and A. M. Hidalgo-Gámez
Departamento de Física, Escuela Superior de Física y Matemáticas, Instituto Politécnico Nacional, U.P. Adolfo López Mateos, Edificio 9, Zacatenco, CP 07730, Mexico City, Mexico; amhidalgog@ipn.mx

## Abstract

We spectroscopically analyzed a sample of 85 Sloan Digital Sky Survey compact, oxygen-dominated galaxies located at redshift $z \sim 0.001$–$0.350$, selected because of their large equivalent width of [O III]$\lambda$5007 (larger than 200 Å). These galaxies might be considered as extreme emission-line galaxies due to their strong [O III]$\lambda$5007 emission line. We detected high-ionization lines, even those related to the presence of Wolf–Rayet stars, in almost all the galaxies studied. We obtained physical properties (electron density and temperature) as well as chemical abundances by using the direct method based on the electron temperature. In this analysis, we obtained three different measurements of $T(high)$: the usual one, $T([O\ III])$, but also that of $T([S\ III])$ and $T([Ar\ III])$ for five and three of the galaxies, respectively. Further, we established a new calibration for $T([S\ III])$. We determined oxygen, nitrogen, sulfur, neon, argon, iron, and chlorine abundances when possible, and we compared to the results of other late-type, low-metallicity galaxies, such as blue compact dwarfs, Ims, and green peas. From such a detailed study, we can conclude that the majority of the galaxies in this sample have similar metallicities to the SMC (about $12 + \log(O/H) \approx 8$ dex), and that only 12% of the galaxies are extremely metal-poor, with abundances lower than 7.7 dex. Also, a comparison with some chemical evolution models as well as a brief discussion on the chemical evolution with time is considered.

*Unified Astronomy Thesaurus concepts:* Metallicity (1031); Chemical abundances (224); Abundance ratios (11); Spectroscopy (1558)

*Supporting material:* machine-readable tables

## 1. Introduction

Emission-line galaxies (ELGs) are very common in the Universe and very useful. These galaxies have a considerable amount of gas (and dust), so significant events of star formation are going on inside them. In those events, massive, hot stars ionize the surrounding gas, and we can detect them up to very large redshift. From their emission-line spectra, we can determine their chemical compositison, but also their rotational velocity and their star formation rates (SFRs), among many other properties (Cohn et al. 2018). Normally, the most intense line in an ELG in the optical part of the spectrum is H$\alpha$ (e.g., Osterbrock & Ferland 2006), with the forbidden oxygen lines at 5007, 4959 Å, and the single oxygen ionized doublet at 3727 Å also quite prominent in the spectra, especially if the galaxies have undersolar abundances (Veilleux & Osterbrock 1987).

However, for some of these ELGs, the most intense line is not H$\alpha$ but [O III]$\lambda$5007, while the intensity of [O II]$\lambda$3727 is quite similar compared to other ELGs. An example of such galaxies is shown in Figure 1. These are referred to as extreme emission-line galaxies (EELGs). Those EELGs have not only the intensity of the [O III]$\lambda$5007 line very large but also its equivalent width (EW), being larger than 100 Å (e.g., Tran et al. 2020), when the normal values are of 50 Å (see Figure 9 in Cardamone et al. 2009; and Figure 2). As the intensity of the oxygen line is the largest of all the lines in the optical part of the spectrum, we prefer to call them *oxygen-dominated* galaxies.

The first to be noticed were some galaxies in the Cardamone et al. (2009) sample of the so-called "green pea" galaxies, and in a sample from the Great Observatories Origins Deep-Survey galaxies (Straughn et al. 2009). Atek et al. (2011) identified some EELGs at $z > 1$ from the WISP survey, and Maseda et al. (2014) studied a sample of 37 galaxies at intermediate–high redshift and concluded that these EELGs are low mass, low metallicity, and can form stars very efficiently. Similar conclusions have been obtained for some other investigations of this type of galaxies, mainly at intermediate to high redshift (e.g., van der Wel et al. 2011; Smit et al. 2014; among others).

There are two simple explanations for the large values of the EW of [O III]$\lambda$5007. The first one is that these galaxies are active galactic nuclei (AGNs); the large intensity of the [O III] $\lambda$5007 being itself a sign of activity. However, most of these EELGs are not classified as AGNs based on several diagnostics (Tang et al. 2021). The second explanation is that these galaxies have very low oxygen abundances, as the intensity of [O III]$\lambda$5007 increases as the oxygen abundances decreases (Campbell 1988). The main caveat is that not many values of the metallicity determinations in EELGs at intermediate or high redshifts have been done so far with reliable methods, such as the electron temperature one (Aller 1984; Osterbrock & Ferland 2006). Although semiempirical methods are good enough for a rough metallicity determination (see Hidalgo-Gámez & Ramírez-Fuentes 2009; Chávez et al. 2014) and very valuable for high-redshift galaxies where few emission lines are detected, for such an important conclusion, a more reliable method is needed. Moreover, it is very well known that all the semiempirical methods, including those based on photoionization models, overestimated the chemical abundances up to 0.2 dex (Hidalgo-Gámez & Ramírez-Fuentes 2009; López-Sánchez & Esteban 2010a; López-Sánchez et al. 2012), although see Pérez-Montero et al. (2021) for a different conclusion based on deep data sets. As said, the bulk of the metallicity determinations in EELGs have been done with semiempirical methods, and very few have been done with the direct one (Amorín et al. 2012; Ly et al. 2014; Amorín et al. 2015). In all of them, the percentage of extremely low-metallicity (hereafter XMP) galaxies is small, between 4% and 15%. Moreover, in most of the previous investigations on EELGs, only the oxygen abundance is determined but not for the rest of the elements of interest (sulfur, nitrogen, neon, and argon are commonly detected in ELGs).

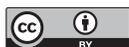

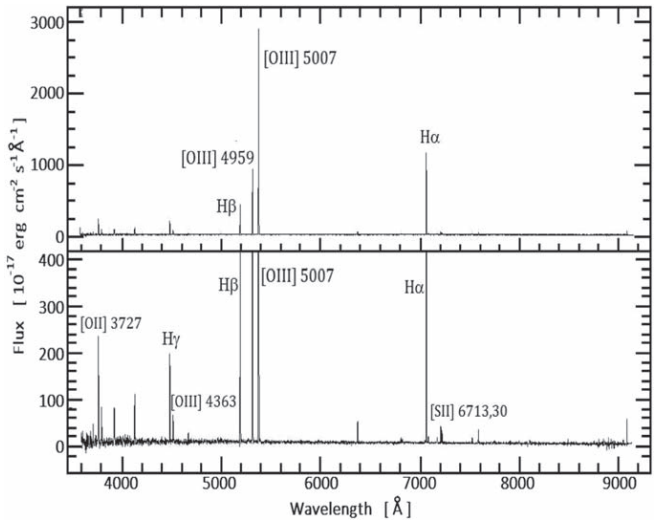

**Figure 1.** The typical optical spectrum of an oxygen-dominated galaxy. It is clear that the most intense line in the spectra is the [O III]λ5007, instead of Hα. The most representative lines are marked with their wavelength and ion. The top panel is the processed SDSS spectrum, and the bottom panel is the same one but with zoom, in order to observe the emission lines in detail.

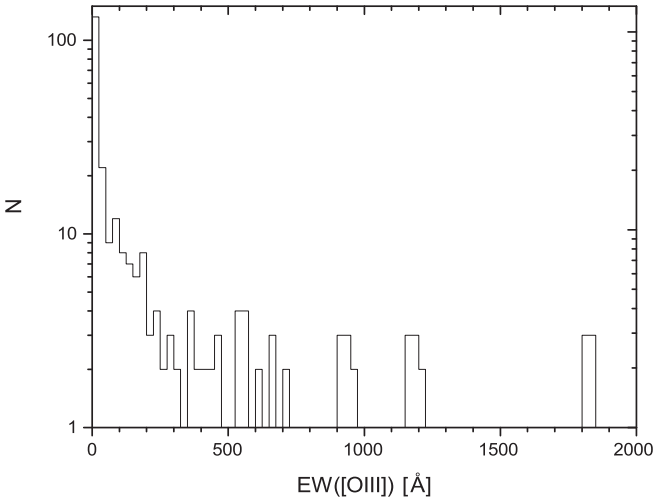

**Figure 2.** Equivalent width of the oxygen line at 5007 Å for a total of 258 late-type galaxies (see text for references). Two breaking points, at 50 and 200 Å, are clearly detected.

In this investigation, we want to explore the chemical abundance of a sample of EELGs using the electron temperature method, in order to check if their high EW([O III]λ5007) and intensities of this line are due to a very low metallicity or rule out this hypothesis. In addition to oxygen, chemical abundances of nitrogen, neon, sulfur, and argon were determined, and, for some galaxies, iron and chlorine were determined as well. Being very few in the determinations of the chemical abundances of the latter elements at high-redshift (Atek et al. 2011; Maseda et al. 2014) and intermediate-redshift EELGs (e.g., Guseva et al. 2020), we think that a bulk of new measurements of the abundances of them might help in the studying of the connection between the large EW of the [O III]λ5007 and the low metal abundance in EELGs.

With those values it can be concluded if the large EW of [O III]λ5007 is due to their low metallicity. If not, there must be another explanation for the large EW and intensity of the [O III] λ5007 line. Steidel et al. (2014) explained the location a sample of 251 galaxies at high $z$, some of them being EELGs, at the BTP diagram by a combination of a hard spectrum in addition to a high-ionization parameter and, possibly, a high density.

The structure of this paper is as follows. We discuss the selection of the sample in the next Section, while the acquisition of the data and its analysis is presented in Section 3. The physical and chemical properties of the galaxies in the sample are studied in Sections 4 and 5, respectively, while a comparison with other ELGs and models as well as a discussion on the reason behind the large EW([O III]) is done in Section 6. Finally, we summarize our conclusion in Section 7.

## 2. Sample Selection

As said, in this investigation, we are going to focus only on those galaxies with an intense (more than usual) EW([O III]) line. However, there is not a clear definition for them. In most of the previous investigations of the EELGs, the lowest limit in the EW([O III]) was typically between 100 and 200 Å. In order to provide a proper definition, we obtained the distribution of the EW of [O III] for a sample of a total of 258 emission-line, star-forming late-type galaxies. They are shown in Figure 2. The total number of galaxies is not very large because the EW([O III]) is very seldom presented in the published investigations. Values from Wolf–Rayet (WR) galaxies (López-Sánchez & Esteban 2010b), hot spots in normal galaxies (Kniazev et al. 2004), and blue compact dwarf (BCD), Im, Sm, and starburst galaxies are included in this figure. The EW([O III]) distribution shows that most of the ELGs have a low value of this parameter. However, there are two *breaking points* in the distribution: one at 50 Å, and the second at 200 Å. The former might correspond to the lower limits of those galaxies with low–normal SFRs. The latter separates the normal ELGs and the EELGs. For EW([O III]) larger than 200 Å, the distribution is very flat, but only 14% of the galaxies in this figure have such values. A similar distribution has been presented in Figure 13 in Cardamone et al. (2009, hereafter C09) for their green peas sample, along with some UVLG and a comparison sample. They concluded that the latter has a very different distribution compared to that of the green peas, but they did not point out any lower limit for the EW([O III]) of their selected galaxies. Actually, the green peas have a broad distribution of EW([O III]) with the low limit of 12.6 Å. In any case, from Figure 2, we can get a first definition of the EELGs as those with EW([O III]) larger than 200 Å. This is the only requirement for a galaxy in order to be considered EELGs.

Now that the definition of the galaxies we are interested in is clear, we can start the real selection of our sample. As we are interested in the obtained reliable chemical abundances not only of oxygen but of other elements, we need high signal-to-noise ratio (S/N) spectra. In those spectra, the oxygen auroral line ([O III]λ4363) might be easily detected, but also lines (nebular mainly) of nitrogen, sulfur, neon, and argon could be detected with enough S/N to obtained a good abundance determination. Moreover, the galaxies should be nearby. The reason is twofold: nearby galaxies are more likely to have large S/N spectra in the Sloan Digital Sky Survey (SDSS), and for galaxies more distant than 0.35, the [S II] lines are quite into the red part of the spectrum, which is noisy, so the density became very difficult to determine. Therefore, the sample was limited to $0 < z < 0.35$.

The SDSS database (Data Release 7, hereafter DR7; York et al. 2000) gives a total of 1398 galaxies with these two conditions. However, only 1043 have an S/N larger than 5 in the auroral oxygen line [O III]λ4363, which is a critical line for a good determination of chemical abundance. Smaller S/N in this line might produce large uncertainties in the abundances, which is not desirable. This sample of 1043 is our master sample (or *sample (1)*) of EELGs or oxygen-dominated galaxies, and it will be used in further investigations. It might be said that such sample is biased because galaxies that have shown the auroral oxygen line [O III]λ4363 have different characteristics than those without it (see, e.g., Hoyos & Díaz 2006). Actually, the galaxies with such a line in the spectra display a very recent event of star formation, younger than 5 Myr according to Martín-Manjón et al. (2008), Villaverde et al. (2010). Any other restriction we might impose to the sample (angular sizes or composite color) has a minor influence on the properties of the galaxies selected compared to this restriction. However, as we are interested in the direct determination of the abundances of other elements besides oxygen, such condition is mandatory in order to determine electron temperatures.

As we are going to determine chemical abundances from fiber spectra of galaxies at different redshifts, we must be sure that those spectra cover the whole galaxies at all distances. This is very important, because it is well known spiral galaxies and some irregular ones have gradients in their metallicity or, at least, important differences in the abundance among different parts of the galaxy (e.g., Vila-Costas & Edmunds 1992). So, we cannot compare abundances from different parts of galaxies without being completely sure that there is no such metallicity variation. Therefore, we selected a subsample that includes galaxies with Petrossian radius between 4″ and 2″, according to their different redshifts (Conselice et al. 2000). This sample, called *sample (2)*, consists of 324 galaxies.

Furthermore, we noticed that some of the galaxies from sample (2) have unusual *gri* composite colors: pink, brown, gray; but mostly green, purple, and bright blue. In appearance, some of the latter resemble the green peas by C09 and the Blueberry Galaxies by Yang et al. (2017). These two new types of galaxies have been selected based on their *gri* composite color, mainly. Although this *gri* color is a false color and might reflect only differences in the redshift, as indicated by Izotov et al. (2011). As a first step in our investigation, we decided to consider only those galaxies with a single color and enough number of data.

Then, our final sample consists of a total of 88 galaxies from sample (2) that have these *gri* colors, and these constitute our *sample (3)*. This sample of 88 objects consists of 23 blue galaxies, 29 purple ones, and 36 green galaxies. Two galaxies of the sample had problems in software compatibility when trying to read the .fits format from SDSS database (J155957.39+004741.0 and J162410.11−002202.6). Another galaxy had a very abnormal red spectrum with very low intensities lines. We decided not to use them. Therefore, the final sample consisted of 85 galaxies.

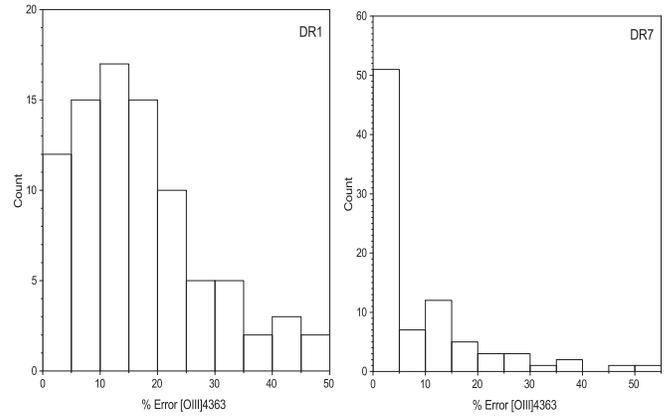

**Figure 3.** Uncertainties, in percentage, in the forbidden oxygen line [O III] λ4363 intensity from SDSS Data Releases, DR1 used by Kniazev et al. (2004), and DR7 (this work).

Although this is not as large as other samples of ELGs or luminous compact ELGs (e.g., Izotov et al. 2021), we must remember that EELGs are less common than normal ELGs, less than 20% according to our Figure 2 and Figure 13 of C09. Concerning the completeness of the sample, none of the samples in this investigation are completed by any means. We are not interested in getting a complete sample but an homogeneous one, as well as a manageable one with statistical significance.

### 2.1. Reliability of the Data

Being public data, it is expected that some of the galaxies have been already studied. We noticed that all the galaxies in our sample (3) were included in an early investigation on luminous galaxies by Kniazev et al. (2004). However, they only determined oxygen abundances, while we are interested in all the chemical abundances that can be estimated. Moreover, they used the Early Data Release (EDR), which is less accurate than the data from the latter releases. This is easy to check from the uncertainties in the auroral oxygen line [O III]λ4363 published by Kniazev et al. (2004). We determined the uncertainties in this line for those galaxies in our sample from the EDR by Kniazev et al. (2004) and compared with the uncertainties in our data by the DR7. This is shown in Figure 3. It is clear that the majority of the galaxies from the DR7 (69%) has uncertainties smaller than 10% while only a third of the galaxies from the EDR have such small uncertainties. Therefore, the chemical abundances determined from EDRs are less accurate, as expected.

We also determined the S/N for this line to check how reliable our abundance determinations were. The S/N of the spectra of our sample is larger than that for previous releases. Actually, 26 of them have S/N larger than 15, and also there are few spectra (five out of 85) with S/N as large as 60.

Finally, although there are few galaxies that have chemical abundance determined by other authors, only two have the abundances of all the chemical elements we presented in this investigation. Therefore, we kept all the galaxies in sample (3) and determined their chemical abundances in a more broad context as EELGs. For those with previous values, a comparison will be given in Section 6.

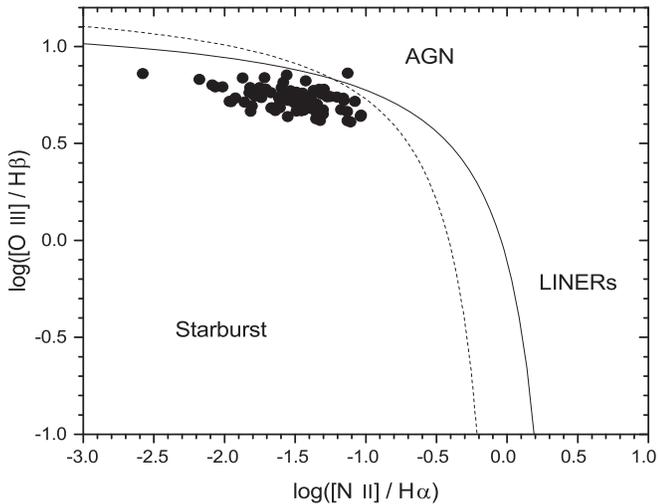

**Figure 4.** The BPT diagram of the galaxies in our sample (3). All of them are classified as ELGs except for one as AGN (the one data point above the lines). The solid line is the curve from Kewley et al. (2001), indicating the upper limit for starburst contribution, and the dashed line corresponds to the model curve from Kauffmann et al. (2003), which separates star-forming objects from AGNs.

### 2.2. Extreme Emission-line Galaxies as Active Galactic Nuclei?

As previously said, the EW of the oxygen line is large in AGNs, even in those of low intensity, like LINERS (Ho et al. 1997). Actually, in many surveys where this line is used to detect ELGs, a certain number of AGNs are "polluting" the ELG sample (e.g., Terlevich et al. 1991). Therefore, we have to exclude all those galaxies that might host an AGN. As not all the AGN galaxies are classified in the catalogs, especially those of low intensity, which might have escaped the attention of the large surveys, we used several ways to check if there is an AGN inside these galaxies. First, we revised the Baldwin–Phillips–Terlevich (BPT) diagram (Baldwin et al. 1981), the one that most clearly separates between AGNs and ELGs (Kauffmann et al. 2003; Groves et al. 2006; Stasińska et al. 2006; Cid Fernandes et al. 2010). This is shown in Figure 4 for the galaxies in our sample (3), along with the model curves that separate AGN regions from a starburst and ELG. All but one galaxy (J004529.15+133908.7) are in the ELG locus, at the very high excitation part of the diagram, which is normal due to the large intensity of the oxygen line. This galaxy is among the top five of galaxies with the largest values of the EW([O III]). However, the other four are well inside the starburst locus.

In addition, by using Table 5 from Ho et al. (1997), where they listed line ratios of several AGNs (LINERS, Seyfert, and transition objects included), we notice all our galaxies fit the H II regions' ones only, including J004529.15+133908.7. This can be clearly seen in Figure 4.

Although the oxygen-dominated galaxies of sample (3) do not have broad lines and their appearance is very different from an AGN spectrum, we can check the FWHM of Hα and [O III] λ5007 for the galaxies in our sample. The average values are 176 km s$^{-1}$ (with a standard deviation of 29 km s$^{-1}$) and 234 km s$^{-1}$ (62 km s$^{-1}$), respectively. Such values are much smaller than the typical values for AGNs, of about 1000 km s$^{-1}$ (Kuraszkiewicz et al. 2004), but also much smaller than the typical FWHM of LINERs, of about 550 km s$^{-1}$ (Torres-Papaqui et al. 2012). Moreover, there is no correlation between the EW([O III]) and the FWHM for any of both lines.

Recently, Schmidt et al. (2021) studied the spectral characteristics of the [O III]λ5007 emission line for Seyfert 1 galaxies. They noticed blue wings in this line with average FWHM of 847 km s$^{-1}$, while it is only of 347 km s$^{-1}$ at the center. These are, again, higher values than those obtained for the galaxies in our sample. Moreover, none of the spectra has the wide blue wing of [O III]λ5007. None of the four tests we performed gave evidence that any of the galaxies we selected were hosting an AGN, even of very low intensity.

### 3. Analysis of the Data

The spectra of the galaxies in the sample were retrieved from SDSS DR7 (York et al. 2000), which were taken using two double fiber-fed spectrometers. Those spectra were downloaded and analyzed by the authors with the ESO-Munich Image Data Analysis System (MIDAS)[1] software, using the ALICE routine through a Gaussian fitting to the line profile. The EW of the line as well as the central wavelength and the intensity of the continuum are provided for each line. With all this information, the total intensity of each line as well as an uncertainty and an S/N can be determined. Such uncertainties were determined from the following:

$$\sigma = \left(\sum_i \sigma_i^2\right)^{1/2}, \quad (1)$$

where $\sigma_i$ is the Poisson noise, and the extinction correction uncertainty ($\sigma_e = \sigma_\alpha^2 + \sigma_\beta^2$).

As the galaxies in the sample cover a large range in redshift, from 0.001 to 0.350, different spectral lines are detected in each spectra. As the wavelength coverage of the SDSS starts at 3800 Å, the lines bluer than this wavelength cannot be detected for the closest galaxies, while the lines at the reddest part of the spectra are outside the range for the farther ones. Therefore, the [O II]λ3727 doublet is not possible to be measured for seven galaxies, while the [S III]λ9069 cannot be detected in galaxies with redshift higher than $z = 0.013$. The oxygen problem can be avoided following Hidalgo-Gámez et al. (2012). The intensity of [O II] can be obtained from the intensity of [O III] from their Equation (2), and the total oxygen abundances can be determined for all the galaxies in the sample. However, the lack of both of nebular lines of sulfur prevents the determination of the sulfur abundances for most of the distant galaxies.

### 3.1. Extinction and Absorption Corrections

Before the analysis of the intensities of the lines, some corrections should be performed. The most important ones being the extinction and stellar absorption corrections. In this investigation, both corrections were performed altogether following López-Sánchez et al. (2006), which, in the last 20 yr, is the best expression that gets both corrections, with the

---
[1] European Southern Observatory-Munich Image Data Analysis System is developed and maintained by the European Southern Observatory.

extinction coefficient determined as

$$c(\mathrm{H}\beta) = \frac{1}{f(\lambda)} \log \left[ \frac{\frac{I(\lambda)}{I(\mathrm{H}\beta)}\left(1 + \frac{W_{\mathrm{H}\beta\mathrm{abs}}}{W_{\mathrm{H}\beta}}\right)}{\frac{F(\lambda)}{F(\mathrm{H}\beta)}\left(1 + \frac{W_{\lambda\mathrm{abs}}}{W_{\lambda}}\right)} \right]. \quad (2)$$

The ratio $F(\lambda)/F(\mathrm{H}\beta)$ is the observed flux ratio, $I(\lambda)/I(\mathrm{H}\beta)$ is the theoretical ratio, $W_{\mathrm{H}\beta\mathrm{abs}}$ and $W_{\lambda\mathrm{abs}}$ are the EWs of the stellar absorption of H$\beta$ and the considered Balmer line, respectively, $W_\lambda$ and $W_{\mathrm{H}\beta}$ are the emission EWs of the considered Balmer line and H$\beta$, respectively, and $f(\lambda)$ is the reddening function. The values for the absorption, $W_{\mathrm{H}\beta\mathrm{abs}}$, $W_{\lambda\mathrm{abs}}$, depend on the starburst age and were taken from Olofsson (1995), while the theoretical line ratios were obtained from Osterbrock & Ferland (2006), and thea reddening function, $f(\lambda)$, values were taken from Toribio San Cipriano et al. (2016). We determined the coefficient with the ratio of H$\alpha$ and H$\beta$ and making comparisons with the rest of the brightest Balmer lines (H$\gamma$ and H$\delta$) until we had the best fits on the observed to theoretical line ratios. The rest of the high-order Balmer lines were not considered on the c(H$\beta$) determination. Once the c(H$\beta$) was obtained, the rest of fluxes of spectral line ratios were corrected with the equation

$$\frac{I(\lambda)}{I(\mathrm{H}\beta)} = \frac{F(\lambda)}{F(\mathrm{H}\beta)} 10^{c(\mathrm{H}\beta)[f(\lambda)]}. \quad (3)$$

We obtained the reddening-corrected intensities of all the lines detected for the 85 galaxies in the sample. However, they are not shown here due to the length of such tables. The authors encourage the readers to contact them to get this information.

It is interesting to notice that the values of the extinction coefficients are quite low, where half of the sample has values between 0 and 0.05. However, although small, such values are not much different from other star-forming galaxies (Masegosa et al. 1994; Toribio San Cipriano et al. 2016) as well as from the ones determined by Kniazev et al. (2004), although their uncertainties in the c(H$\beta$) values are higher due to their large uncertainties.

According to some authors (e.g., Scarlata 2019), those low (or negative) values of the c(H$\beta$) cannot be fitted with any of the classical recombination cases (Brocklehurst 1971) but a brand new one. However, we noticed that these negative values are due to the use of the simplified c(H$\beta$) definition based only in the intensities of the H$\alpha$ and H$\beta$ lines. But when a most-elaborated definition is considered, as that shown in Equation (2), the c(H$\beta$) values are positive.

Line intensities, extinction and absorption corrected, can be found in the Appendix section.

## 4. Physical Properties of the Oxygen-dominated Galaxies

The main goal of this investigation is to determine the chemical abundances of the oxygen-dominated galaxies in our sample, with a focus on other elements than oxygen. Although there is information on the oxygen abundance for several EELGs, the majority have no information about the other elements but oxygen. Therefore, we studied a representative sample (85 galaxies) with S/N larger than 5 in [O III]$\lambda$4363, which allows the determinations of chemical abundances using the reliable direct method.

### 4.1. Electron Density

The electron density was determined from the sulfur line ratio [S II]$\lambda\lambda$6717, 6731 following Castañeda et al. (1992). For ten galaxies, the canonical value of $n_e = 100 \, \mathrm{cm}^{-3}$ was considered because the ratio of the sulfur lines was higher than 1.45. For another 13 galaxies, the sulfur lines were not resolved, and an upper limit of $50 \, \mathrm{cm}^{-3}$ was set. Then, the density was computed for a total of 62 galaxies, and the final value range is very wide, from low densities ($n_e \approx 60 \, \mathrm{cm}^{-3}$) to very high values ($>1000 \, \mathrm{cm}^{-3}$). The average density is $300 \, \mathrm{cm}^{-3}$, which is very similar to the values obtained by Amorín et al. (2012) for a small sample of green peas.

There might be several reasons for such large density values. It is well known that the existence of shocks increases the [S II] intensity (Jaskot & Oey 2013). As previously said, shocks are not an important source of ionization for the galaxies in our sample. Moreover, for our sample with the largest densities, the lines associated to shocks (as [O I]$\lambda$6300, [N I]$\lambda$5200, or [O I]$\lambda$6363) were detected in nine, two, and three galaxies only with faint intensities. Therefore, we do not think the high density obtained is because of shocks.

Another possible reason might be the existence of very dense clumps of gas and dust inside the galaxy, as those presented in the H II regions in our Galaxy (Vilchez & Esteban 1996; Copetti et al. 2000; Heyer et al. 2016). They might increase the total density. However, it is very difficult to explain how very small clumps with sizes of 1 pc (or less) dominate the density at kiloparsec scales. This could be due to the integrated optical spectra, such as those studied here, being dominated by the densest regions in the galaxies (H. Castañeda 2020, private communication).

Finally, we can check a few ideas on the reasons for such large densities. One might think that larges densities might be due to a large amount of gas inside these galaxies. Such gas is used for star formation. Therefore, the larger the density, the larger the SFR (see Kaasinen et al. 2017 for high-$z$ galaxies; and Kashino & Inoue 2019 for low-$z$ ones). However, there are no relations between the SFR and the density. Not all the galaxies with the largest SFRs are the densest, or vice versa.

It might be interesting to check the density values with other density diagnostic lines. Unfortunately, our spectra does not contain any other $n_e$ tracer. So, higher quality spectra are needed to establish the origin of the high electron density values for the EELGs.

### 4.2. Electron Temperatures

As previously stated, the forbidden auroral line of oxygen at 4363 Å is detected in all the spectra. Therefore, the electron temperature could be determined, and the chemical abundances can be obtained by the so-called direct method (Aller 1984; Osterbrock & Ferland 2006).

Also, the sulfur line at [S III]$\lambda$9069 was detected in five galaxies, so along with the line at [S III]$\lambda$6312, the electron temperature for this ion was determined. Another similar situation occurs with the auroral argon line, [Ar III]$\lambda$5192, detected in three objects, so along with the line [Ar III]$\lambda$7135, it is possible to compute the electron temperature of this ion in these objects. Finally, other electron temperatures, such as those of [S II] and [O II], could be determined for four and eight galaxies, respectively. Henceforward, we describe all these temperature determinations and how they relate with each other.

#### 4.2.1. High-ionization Temperatures

The most common element for obtaining the temperature of the so-called high-ionization zone is the oxygen because both its nebular and auroral lines are less than 1000 Å apart, and its

**Table 1**
Physical Properties of the Different Elements Obtained in Our Sample of Oxygen-dominated Galaxies

| ID SDSS | $n_e$([S III]) (cm$^{-3}$) | $t_e$([O III]) ($\times 10^4$) K | $t_e$([S III]) ($\times 10^4$) K | $t_e$([S III])$_{tw}$ ($\times 10^4$) K | $t_e$([Ar III]) ($\times 10^4$) K | $t_e$([S II]) ($\times 10^4$) K | $t_e$([O II]) ($\times 10^4$) K | $t_e$([O II])$_{I06}$ ($\times 10^4$) K |
|---|---|---|---|---|---|---|---|---|
| J003218.60+150014.2 | <50 | 1.41 ± 0.03 | ⋯ | 1.47 ± 0.16 | ⋯ | ⋯ | ⋯ | 1.58 ± 0.02 |
| J004054.33+153409.7 | 580 ± 10 | 1.37 ± 0.00 | ⋯ | 1.44 ± 0.18 | ⋯ | ⋯ | ⋯ | 1.50 ± 0.00 |
| J004236.93+160202.7 | 100 ± 41 | 1.21 ± 0.11 | ⋯ | 1.31 ± 0.22 | ⋯ | ⋯ | ⋯ | 1.25 ± 0.09 |
| J004529.15+133908.7 | 1097 ± 474 | 1.59 ± 0.14 | ⋯ | 1.62 ± 0.12 | ⋯ | ⋯ | ⋯ | 1.43 ± 0.05 |
| J005147.30+001940.0 | 100 ± 13 | 1.72 ± 0.02 | ⋯ | 1.73 ± 0.08 | ⋯ | ⋯ | ⋯ | 1.47 ± 0.01 |
| J010513.47−103741.0 | 155 ± 18 | 1.12 ± 0.10 | ⋯ | 1.23 ± 0.24 | ⋯ | ⋯ | 1.42 ± 0.05 | 1.12 ± 0.09 |
| J013344.63+005711.2 | <50 | 1.47 ± 0.01 | ⋯ | 1.52 ± 0.15 | ⋯ | ⋯ | ⋯ | 1.70 ± 0.01 |
| J014721.68−091646.3 | 108 ± 14 | 1.16 ± 0.13 | ⋯ | 1.27 ± 0.23 | ⋯ | ⋯ | ⋯ | 1.17 ± 0.11 |
| J020051.59−084542.9 | <50 | 1.11 ± 0.11 | ⋯ | 1.23 ± 0.24 | ⋯ | ⋯ | ⋯ | 1.12 ± 0.10 |
| J021852.90−091218.7 | 154 ± 15 | 1.99 ± 0.10 | 2.10 ± 0.24 | 1.96 ± 0.01 | ⋯ | ⋯ | ⋯ | 1.49 ± 0.02 |
| J022907.37−085726.2 | 100 ± 27 | 1.15 ± 0.16 | ⋯ | 1.26 ± 0.23 | ⋯ | ⋯ | ⋯ | 1.19 ± 0.34 |
| J024052.20−082827.4 | 242 ± 15 | 1.60 ± 0.05 | ⋯ | 1.63 ± 0.11 | ⋯ | ⋯ | 1.34 ± 0.02 | 1.44 ± 0.02 |
| J024453.67−082137.9 | 100 ± 6 | 1.33 ± 0.05 | ⋯ | 1.41 ± 0.18 | ⋯ | ⋯ | ⋯ | 1.44 ± 0.04 |
| J025346.70−072344.1 | 100 ± 6 | 1.46 ± 0.05 | 1.51 ± 0.09 | 1.52 ± 0.15 | ⋯ | ⋯ | ⋯ | 1.68 ± 0.03 |
| J025436.11−000138.0 | 100 ± 65 | 1.14 ± 0.13 | ⋯ | 1.25 ± 0.24 | ⋯ | ⋯ | ⋯ | 1.17 ± 0.27 |
| J030321.41−075923.2 | 365 ± 9 | 1.69 ± 0.10 | ⋯ | 1.71 ± 0.09 | ⋯ | ⋯ | ⋯ | 1.47 ± 0.02 |
| J030539.71−083905.3 | 130 ± 26 | 1.37 ± 0.26 | ⋯ | 1.44 ± 0.17 | ⋯ | ⋯ | ⋯ | 1.52 ± 0.13 |
| J031623.96+000912.3 | 201 ± 22 | 1.28 ± 0.00 | ⋯ | 1.37 ± 0.20 | ⋯ | ⋯ | ⋯ | 1.35 ± 0.00 |
| J032613.63−063513.5 | 196 ± 4 | 1.32 ± 0.15 | ⋯ | 1.40 ± 0.19 | ⋯ | ⋯ | ⋯ | 1.42 ± 0.09 |
| J033031.22−005846.7 | <50 | 1.21 ± 0.07 | ⋯ | 1.31 ± 0.22 | ⋯ | ⋯ | ⋯ | 1.25 ± 0.06 |
| J033947.79−072541.3 | 107 ± 11 | 1.14 ± 0.07 | ⋯ | 1.25 ± 0.24 | ⋯ | ⋯ | ⋯ | 1.15 ± 0.06 |
| J040937.63−051805.8 | 101 ± 8 | 1.30 ± 0.09 | ⋯ | 1.38 ± 0.19 | 1.36 ± 0.03 | ⋯ | ⋯ | 1.38 ± 0.06 |
| J075315.74+401449.9 | 253 ± 20 | 2.06 ± 0.37 | ⋯ | 2.01 ± 0.01 | ⋯ | ⋯ | ⋯ | 1.48 ± 0.15 |
| J075715.74+452137.9 | 100 ± 1 | 1.28 ± 0.19 | ⋯ | 1.37 ± 0.20 | ⋯ | ⋯ | ⋯ | 1.35 ± 0.12 |
| J080147.11+435302.3 | 142 ± 1 | 1.03 ± 0.08 | ⋯ | 1.16 ± 0.26 | 1.17 ± 0.01 | ⋯ | ⋯ | 1.03 ± 0.10 |
| J082530.68+504804.5 | 119 ± 3 | 1.19 ± 0.05 | ⋯ | 1.30 ± 0.22 | ⋯ | ⋯ | ⋯ | 1.22 ± 0.04 |
| J083350.24+454933.6 | 334 ± 3 | 1.13 ± 0.10 | ⋯ | 1.24 ± 0.24 | ⋯ | ⋯ | ⋯ | 1.13 ± 0.09 |
| J083440.06+480540.9 | 289 ± 7 | 1.19 ± 0.12 | ⋯ | 1.29 ± 0.22 | ⋯ | ⋯ | ⋯ | 1.22 ± 0.10 |
| J084029.91+470710.2 | 416 ± 8 | 2.01 ± 0.03 | ⋯ | 1.97 ± 0.01 | ⋯ | ⋯ | ⋯ | 1.54 ± 0.00 |
| J084527.61+530853.0 | 100 ± 4 | 1.23 ± 0.04 | ⋯ | 1.33 ± 0.21 | ⋯ | ⋯ | ⋯ | 1.28 ± 0.03 |
| J085207.68−001118.0 | 155 ± 4 | 1.07 ± 0.14 | ⋯ | 1.20 ± 0.25 | ⋯ | ⋯ | ⋯ | 1.07 ± 0.24 |
| J090047.44+574255.2 | 77 ± 7 | 0.99 ± 0.14 | ⋯ | 1.12 ± 0.28 | ⋯ | ⋯ | ⋯ | 0.98 ± 0.16 |
| J090122.81−002818.9 | 100 ± 72 | 1.36 ± 0.24 | ⋯ | 1.43 ± 0.18 | ⋯ | ⋯ | ⋯ | 1.49 ± 0.13 |
| J090139.86+575946.2 | 91 ± 3 | 1.13 ± 0.19 | ⋯ | 1.24 ± 0.24 | ⋯ | ⋯ | ⋯ | 1.15 ± 0.39 |
| J091652.24+003113.9 | 167 ± 11 | 0.98 ± 0.08 | ⋯ | 1.12 ± 0.28 | ⋯ | ⋯ | ⋯ | 0.98 ± 0.09 |
| J092635.25+582047.3 | 382 ± 11 | 1.03 ± 0.30 | ⋯ | 1.16 ± 0.27 | ⋯ | ⋯ | ⋯ | 1.02 ± 0.27 |
| J092918.39+002813.2 | 113 ± 2 | 1.34 ± 0.24 | ⋯ | 1.42 ± 0.18 | ⋯ | ⋯ | ⋯ | 1.46 ± 0.13 |
| J093006.60+602653.0 | 100 ± 36 | 1.40 ± 0.03 | 1.72 ± 0.04 | 1.47 ± 0.17 | ⋯ | ⋯ | ⋯ | 1.56 ± 0.02 |
| J094850.92+553716.1 | 153 ± 91 | 1.39 ± 0.26 | ⋯ | 1.45 ± 0.17 | ⋯ | ⋯ | ⋯ | 1.54 ± 0.13 |
| J095023.32+004229.2 | 196 ± 2 | 1.26 ± 0.06 | ⋯ | 1.36 ± 0.20 | ⋯ | ⋯ | ⋯ | 1.33 ± 0.04 |
| J103344.05+635317.3 | 101 ± 42 | 1.18 ± 0.11 | ⋯ | 1.29 ± 0.22 | ⋯ | ⋯ | ⋯ | 1.21 ± 0.09 |
| J104554.78+010405.8 | 91 ± 1 | 1.23 ± 0.02 | ⋯ | 1.33 ± 0.21 | ⋯ | 0.40 ± 0.08 | ⋯ | 1.28 ± 0.02 |
| J105032.49+661654.0 | 130 ± 10 | 1.49 ± 0.08 | ⋯ | 1.54 ± 0.14 | ⋯ | ⋯ | ⋯ | 1.73 ± 0.04 |
| J112502.58−004525.6 | 257 ± 73 | 1.44 ± 0.11 | ⋯ | 1.50 ± 0.16 | ⋯ | ⋯ | ⋯ | 1.64 ± 0.06 |
| J113303.80+651341.3 | 100 ± 57 | 1.31 ± 0.23 | ⋯ | 1.39 ± 0.19 | ⋯ | ⋯ | ⋯ | 1.41 ± 0.14 |
| J113341.19+634925.9 | <50 | 1.36 ± 0.08 | 1.51 ± 0.08 | 1.44 ± 0.18 | ⋯ | ⋯ | ⋯ | 1.50 ± 0.05 |
| J113459.60−000104.2 | 499 ± 177 | 1.28 ± 0.28 | ⋯ | 1.37 ± 0.20 | ⋯ | ⋯ | ⋯ | 1.36 ± 0.17 |
| J114649.34+005346.0 | 154 ± 25 | 1.99 ± 0.08 | ⋯ | 1.96 ± 0.01 | ⋯ | ⋯ | ⋯ | 1.49 ± 0.01 |
| J115247.52−004007.7 | 121 ± 8 | 1.57 ± 0.06 | 1.53 ± 0.11 | 1.60 ± 0.12 | ⋯ | ⋯ | ⋯ | 1.42 ± 0.02 |
| J120055.64+032404.0 | <50 | 1.64 ± 0.04 | ⋯ | 1.66 ± 0.10 | ⋯ | ⋯ | ⋯ | 1.45 ± 0.01 |
| J122419.73+010559.5 | <50 | 1.26 ± 0.10 | ⋯ | 1.35 ± 0.20 | ⋯ | ⋯ | ⋯ | 1.32 ± 0.07 |
| J123436.30−020721.3 | <50 | 1.13 ± 0.10 | ⋯ | 1.24 ± 0.24 | ⋯ | ⋯ | ⋯ | 1.16 ± 0.19 |
| J125526.06−021334.1 | <50 | 1.65 ± 0.07 | ⋯ | 1.67 ± 0.10 | ⋯ | ⋯ | ⋯ | 1.45 ± 0.02 |
| J130029.30+021502.9 | 315 ± 78 | 2.07 ± 0.40 | ⋯ | 2.02 ± 0.01 | ⋯ | ⋯ | ⋯ | 1.54 ± 0.17 |
| J130148.03+013718.6 | 181 ± 3 | 1.12 ± 0.14 | ⋯ | 1.23 ± 0.24 | ⋯ | ⋯ | ⋯ | 1.12 ± 0.13 |
| J130211.15−000516.4 | 330 ± 6 | 0.99 ± 0.25 | ⋯ | 1.13 ± 0.28 | ⋯ | ⋯ | ⋯ | 0.98 ± 0.24 |
| J130249.19+653449.5 | <50 | 1.33 ± 0.06 | ⋯ | 1.41 ± 0.19 | ⋯ | ⋯ | ⋯ | 1.43 ± 0.04 |
| J131654.37−024930.4 | 62 ± 46 | 1.75 ± 0.17 | ⋯ | 1.76 ± 0.07 | ⋯ | 0.52 ± 0.08 | ⋯ | 1.48 ± 0.02 |
| J131937.24+005043.8 | 87 ± 16 | 0.97 ± 0.06 | ⋯ | 1.12 ± 0.28 | ⋯ | ⋯ | ⋯ | 0.98 ± 0.05 |
| J132347.46−013253.0 | 907 ± 67 | 1.84 ± 0.04 | ⋯ | 1.83 ± 0.05 | ⋯ | ⋯ | ⋯ | 1.49 ± 0.00 |
| J132654.62+011346.8 | <50 | 1.32 ± 0.06 | ⋯ | 1.40 ± 0.19 | ⋯ | ⋯ | ⋯ | 1.42 ± 0.04 |
| J133303.96+624603.8 | 93 ± 28 | 2.15 ± 0.56 | ⋯ | 2.08 ± 0.03 | ⋯ | ⋯ | ⋯ | 1.53 ± 0.35 |
| J135155.90+032524.3 | 1062 ± 35 | 1.07 ± 0.12 | ⋯ | 1.20 ± 0.25 | ⋯ | ⋯ | ⋯ | 1.07 ± 0.11 |

**Table 1**
(Continued)

| ID SDSS | $n_e$([S III]) (cm$^{-3}$) | $t_e$([O III]) ($\times 10^4$) K | $t_e$([S III]) ($\times 10^4$) K | $t_e$([S III])$_{tw}$ ($\times 10^4$) K | $t_e$([Ar III]) ($\times 10^4$) K | $t_e$([S II]) ($\times 10^4$) K | $t_e$([O II]) ($\times 10^4$) K | $t_e$([O II])$_{I06}$ ($\times 10^4$) K |
|---|---|---|---|---|---|---|---|---|
| J135930.37−010322.2 | 68 ± 23 | 0.83 ± 0.13 | ⋯ | 1.00 ± 0.32 | ⋯ | ⋯ | ⋯ | 0.93 ± 0.04 |
| J140725.31+052837.9 | 100 ± 15 | 1.35 ± 0.14 | ⋯ | 1.42 ± 0.18 | 1.37 ± 0.09 | ⋯ | ⋯ | 1.47 ± 0.09 |
| J141940.33+050906.8 | 166 ± 9 | 1.29 ± 0.53 | ⋯ | 1.37 ± 0.20 | ⋯ | ⋯ | ⋯ | 1.37 ± 0.23 |
| J142214.31−003919.5 | 190 ± 16 | 1.43 ± 0.18 | ⋯ | 1.49 ± 0.16 | ⋯ | ⋯ | ⋯ | 1.62 ± 0.09 |
| J143804.21+013333.5 | 1371 ± 208 | 1.14 ± 0.15 | ⋯ | 1.25 ± 0.23 | ⋯ | ⋯ | ⋯ | 1.15 ± 0.13 |
| J144205.41−005248.6 | 319 ± 39 | 1.08 ± 0.14 | ⋯ | 1.20 ± 0.25 | ⋯ | ⋯ | ⋯ | 1.07 ± 0.13 |
| J144805.38−011057.7 | 183 ± 4 | 1.32 ± 0.02 | ⋯ | 1.40 ± 0.19 | ⋯ | ⋯ | ⋯ | 1.42 ± 0.01 |
| J152830.72+001740.2 | 148 ± 17 | 1.16 ± 0.17 | ⋯ | 1.27 ± 0.23 | ⋯ | ⋯ | 1.50 ± 0.08 | 1.18 ± 0.14 |
| J154108.38+032029.4 | 123 ± 23 | 1.37 ± 0.18 | ⋯ | 1.44 ± 0.18 | ⋯ | ⋯ | ⋯ | 1.50 ± 0.10 |
| J154337.31−000608.1 | 1160 ± 460 | 1.50 ± 0.33 | ⋯ | 1.55 ± 0.14 | ⋯ | ⋯ | ⋯ | 1.39 ± 0.10 |
| J154654.55+030902.2 | 73 ± 6 | 1.35 ± 0.06 | ⋯ | 1.43 ± 0.18 | ⋯ | 1.84 ± 0.07 | ⋯ | 1.47 ± 0.04 |
| J155644.29+540328.9 | <50 | 1.23 ± 0.18 | ⋯ | 1.33 ± 0.21 | ⋯ | ⋯ | ⋯ | 1.28 ± 0.13 |
| J164359.16+443633.0 | 344 ± 23 | 1.08 ± 0.16 | ⋯ | 1.20 ± 0.25 | ⋯ | ⋯ | ⋯ | 1.08 ± 0.28 |
| J172009.83+542133.2 | 1859 ± 553 | 1.28 ± 0.16 | ⋯ | 1.37 ± 0.20 | ⋯ | ⋯ | ⋯ | 1.36 ± 0.11 |
| J172706.33+594902.2 | 100 ± 58 | 1.25 ± 0.18 | ⋯ | 1.34 ± 0.21 | ⋯ | ⋯ | ⋯ | 1.31 ± 0.13 |
| J173126.55+591150.1 | 412 ± 24 | 1.29 ± 0.22 | ⋯ | 1.38 ± 0.20 | ⋯ | ⋯ | ⋯ | 1.37 ± 0.14 |
| J173501.25+570308.6 | 244 ± 1 | 1.30 ± 0.02 | ⋯ | 1.38 ± 0.19 | ⋯ | ⋯ | ⋯ | 1.38 ± 0.01 |
| J174327.60+544320.1 | 535 ± 27 | 1.22 ± 0.25 | ⋯ | 1.32 ± 0.21 | ⋯ | ⋯ | ⋯ | 1.26 ± 0.17 |
| J210114.40−055510.3 | 143 ± 5 | 1.13 ± 0.08 | ⋯ | 1.24 ± 0.24 | ⋯ | ⋯ | ⋯ | 1.14 ± 0.07 |
| J211527.07−075951.4 | <50 | 1.04 ± 0.07 | ⋯ | 1.17 ± 0.26 | ⋯ | ⋯ | ⋯ | 1.03 ± 0.07 |
| J232936.55−401105.9 | <50 | 1.26 ± 0.08 | ⋯ | 1.35 ± 0.20 | ⋯ | ⋯ | ⋯ | 1.33 ± 0.06 |
| J235347.69+005402.1 | 100 ± 32 | 1.36 ± 0.26 | ⋯ | 1.43 ± 0.18 | ⋯ | ⋯ | ⋯ | 1.49 ± 0.13 |

**Note.** The density is determined following Castañeda et al. (1992). The $t_e$([S III])$_{tw}$ obtained with the relation proposed in this work. The $t_e$([O II])$_{I06}$ calculated following Izotov et al. (2006b). The rest of the temperatures were determined by the direct method (Aller 1984).

(This table is available in machine-readable form.)

auroral line at 4363 Å is the brightest for star-forming galaxies. Moreover, when no other auroral line is detected, this temperature is used to determine the ionic abundances of all the high-ionization species, as [S III], [Ne III] or [Ar IV]. This procedure is widely used (e.g., Amorín et al. 2012; Toribio San Cipriano et al. 2016), and we will discuss its accuracy.

We determined the $t_e$([O III]) following Equation (5.4) from the Osterbrock & Ferland (2006) classic method. The obtained values cover a wide range, from 8300 to 21,500 K, with an average value of 13,400 K. However, six galaxies showed temperatures higher than 19,000 K. They might be considered as extremely metal-poor galaxies (Papaderos et al. 2006; Sánchez Almeida et al. 2016; Annibali et al. 2019), because the higher the temperature, the lower the metallicity (Masegosa et al. 1994). The $t_e$([O III]) determined here are very similar to the values obtained for other ELGs (Pérez-Montero & Díaz 2003, hereafter PM03; Hägele et al. 2006, 2011, 2012; Amorín et al. 2012; Pérez et al. 2016).

The values of the $t_e$([Ar III]), obtained with the ratio between the nebular and auroral lines at λλ7135, 7751, and λ5192 respectively, are very similar to those from $t_e$([O III]), so the abundances obtained based on the [O III] temperatures are accurate enough, given the uncertainties. Therefore, we will use $t_e$([O III]) to determine the argon abundances for all the spectra in our sample except for those three galaxies.

For five galaxies, the intensities of the [S III]λ9069 and [S III] λ6312 lines were measured, so the $t_e$([S III]) could be determined following Aller (1984) formalism. We noticed that the $t_e$([S III]) does not change when new atomic data are used. All the values are higher than 15,000 K, which are slightly higher than the $t_e$([O III]) values for the same galaxies (see Table 1). Therefore, the use of $t_e$([O III]) for the sulfur abundance might be not very accurate. A similar result is obtained by PM03, where seven out of 11 galaxies have $t_e$([S III]) values higher than $t_e$([O III]), and the former is always higher than 15,000 K.

Due to the weakness of the sulfur auroral line and because the nebular ones are located in the far red region of the optical spectra, the $t_e$([S III]) cannot be determined directly for many spectra. In order to obtain this temperature, two different approaches are considered. The first one is to use the $t_e$([O III]) as a generic temperature for all the elements in the high-ionization zone. If $t_e$([S III]) is higher (lower) than $t_e$([O III]), the sulfur abundances determined with the oxygen temperature might be underestimated (overestimated). The second one is to obtain a correlation between the oxygen and sulfur temperatures. There are several of such correlations in the literature, some based on data (Hägele et al. 2006) and others based on photoionization models (e.g., Garnett 1992; or Izotov et al. 2006b). However, we noticed that they are very different, and when low-ionization regions are included, as those in M33 by López-Hernández et al. (2013), the fitting changes significantly from those relationships where only high-ionization regions are considered. With this in mind, we decided to gather $t_e$([S III]) values from high-ionization galaxies and get a relationship from them. The results are shown in Figure 5, where a total of 29 data points are considered: our five spectra, 12 H II galaxies from PM03, the BCD galaxy Haro 15 from Hägele et al. (2012), three compact H II galaxies from Hägele et al. (2006), three H II regions in an SDSS H II galaxy (Hägele et al. 2011), and four H II regions from NGC 5253 (López-Sánchez et al. 2007; LS07). Along with data points, the 1:1 line (dashed line) is shown, as well as the least squares fitting (solid line) to all these data points, which is

$$t(S^{++}) = (0.82 \pm 0.15)t(O^{++}) + (0.314 \pm 0.22). \quad (4)$$

We compared the values we obtained from this fitting to those obtained with the model by Izotov et al. (2006b), and noticed that they are similar only for a narrow range of temperatures, from 11,000 to 15,000 K. Outside of it, the model overestimates (or underestimate) the real sulfur temperatures. We decided to use the $t$([S III]) obtained from our fitting because these values are optimized for high-ionization galaxies. Another reason is the good correlation between the $t$([S III]) and the sulfur abundance, as can be seen in Figure 6.

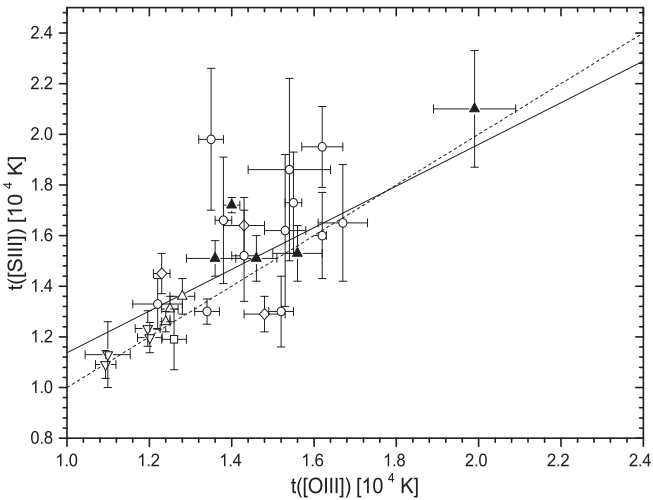

**Figure 5.** Correlation between $t_e$([O III]) and $t_e$([S III]) for the data analyzed in this work (filled triangles), and the data points taken from the literature (circle, PM03; open triangle, Hägele et al. 2006; inverted open triangle, LS07; diamond, Hägele et al. 2011; and square, Hägele et al. 2012; see text for references). Along with data points, the fitting curve (solid line) and 1:1 line (dashed line) are shown.

Figure 7 shows the distribution of oxygen and sulfur temperatures, along with the average value for our sample (dashed lines), three green peas galaxies (Amorín et al. 2012), some XMP galaxies (Papaderos et al. 2006), and some H II galaxies (Hägele et al. 2006, 2008). Both distributions are broad with a long *tail* toward high temperatures. Our values are very similar to those of the GP galaxies and to those of the H II galaxies, but not to those of XMP for both temperatures. Therefore, the values of the electron temperature of the oxygen-dominated galaxies are very similar to those from other [O III]-dominated galaxies, and the main difference might be the slightly larger number of galaxies with high $t$([O III]) than in other samples.

#### 4.2.2. Low-ionization Temperatures

The low-ionization temperatures are more difficult to determined because the lines involved are weaker than those of the high-ionization zone. Although, due to the high S/N of the spectra used in this work, we can determine the $T$([O II]) for five spectra and the $T$([S II]) for another three. However, only three and one values respectively are reliable because they are in the expected range. We will use those temperatures for the determination of the ionic abundances of $O^+$ and $S^+$ for these galaxies. For the rest of the sample, we used expression (14) from Izotov et al. (2006b) to determine the oxygen temperature, and considered that $T(O^+) = T(S^+) = T(N^+)$, as usual. The values are ranging from 9500 to 16,500 K for the galaxies in the sample, being typical values of ELGs (Izotov et al. 2006a; Guseva et al. 2007). We show the electron temperatures of oxygen-dominated galaxies in Table 1.

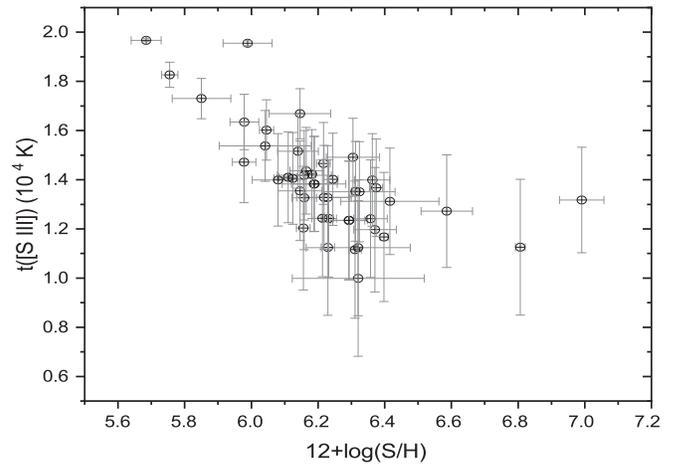

**Figure 6.** Correlation between the electron temperature and the abundance for sulfur for the data of our sample. The $t_e$([S III]) was determined following Equation (4), and the sulfur abundance was determined with this temperature.

### 5. Chemical Abundances

As said, the goal of this investigation is to determine the chemical abundances of the EELGs for other elements besides oxygen and explore if the large intensities and EW of the oxygen lines have any consequence on the chemical abundances. So the abundances' values of sulfur, nitrogen, argon, and neon, as well as oxygen, were determined. In addition, lines of chlorine and iron were also detected, and their abundances were determined for some of the galaxies in the sample.

The ionic abundances of all the elements detected were determined following Equations (4)–(5) of Aller (1984) for the $p^2$ and $p^4$ configuration ions, such as $O^{++}$, $S^{++}$, $N^+$, $Ne^{++}$, and $Ar^{++}$; while we used their Equation (5) for the $p^3$ configuration ions ($O^+$, $S^+$, $Ar^{3+}$), except for iron and chlorine where we used expressions (10) and (13) of Izotov et al. (2006b). We also used new atomic parameters, but the differences in the abundance determinations were smaller than the uncertainties. A similar conclusion is reached by Izotov et al. (2006b).

The atomic abundances determination was performed with the aid of the so-called ionization correction factor (hereafter ICF). Nowadays, there are several ICFs for each element, so it is needed to check which one provides the best results for our sample. For nitrogen and argon, different ICFs (Torres-Peimbert et al. 1989; Izotov et al. 2006b; Pérez-Montero et al. 2007) provide consistent abundances within errors. For the rest of the elements, different ICFs give different atomic abundances. We used the log(X/O) versus 12 + log(O/H) plot (see next subsection) to decide which the most appropriate ICFs are for each atom. We decided to use the one by Torres-Peimbert et al. (1989) for nitrogen, Izotov et al. (2006b) for sulfur, Torres-Peimbert et al. (1989) for neon, and Pérez-Montero et al. (2007) for argon. Finally, the iron ICF described by Rodríguez & Rubin (2005) gives a smaller dispersion in the log(X/O) versus 12 + log(O/H) plot, obtained only for 12 galaxies. For the five galaxies where it is possible to determine chlorine abundances, the ICF from Izotov et al. (2006b) is used.

The total abundances of all these chemical species (oxygen, nitrogen, sulfur, neon, argon, chlorine, and iron) are listed in Table 2 and can be summarized in Figure 8. The uncertainties of the total abundances were determined from the uncertainties

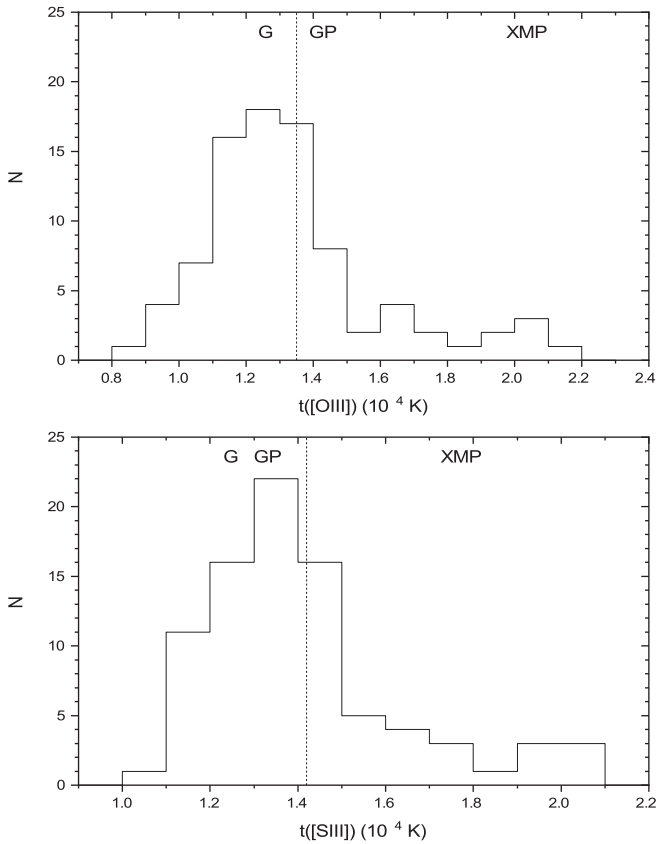

**Figure 7.** Distribution of the electron temperature of $O^{++}$ and $S^{++}$. The average value for our sample is indicated with the dashed line, while "GP" stands for green peas, "XMP" stands for extremely metal-poor galaxies, and "G" stands for H II galaxies (see text for references).

in the line intensities, and those in the extinction correction by quadrature (Hidalgo-Gámez 2012).

The distribution of the oxygen abundance covers from 7.46 to 8.34 dex, as seen in Figure 8(a), all of them being subsolar abundance, with an average oxygen abundance of 8.00 dex. The oxygen abundance is very similar to those values from other kinds of ELG, although there is high intensity for the oxygen nebular lines (see Section 6.1).

Sulfur abundances are listed in column (4) of Table 2. We could not estimate it for all the sampled galaxies because of the absence of, at least, one of the ions involved ($S^+$ or $S^{2+}$). For all but five galaxies, the $t_e$ used in the abundance determination was determined from relation in Equation (4). The temperature obtained with this equation is accurate enough as the differences are within the uncertainties. The values of sulfur abundance range from $12 + \log(S/H) = 5.68$ to 6.99 dex (Figure 8(b)) with an average value of 6.21 dex, being all subsolar and typical ones for H II galaxies (Pagel 2009). However, these values shift toward smaller abundances than other ELGs (see Table 4). Such low sulfur abundances are not due to to the $t_e(S^{++})$ we used (any other temperature increases the abundances less than 0.1 dex), or to any other ICF (all the other ICFs give similar abundances' values).

Neon is a very common element in the H II regions of late-type galaxies. We could determine the neon abundances for all the galaxies in our sample except three (J025436.11-000138.0, J083440.06+480540.9, and J155644.29+540328.9) for which no neon line was detected. The values are ranging from $12 + \log(Ne/H) = 6.76$ to 7.87 dex (see Figure 8(c)). The average neon abundance is 7.39 dex. Although this value is similar to other ELGs (see Section 6.2 for a discussion), the range is quite broad, with an important number of galaxies with low neon abundances.

The average value for nitrogen is 6.35 dex, which is very low compared to other ELGs (see Table 4). However, the distribution is broad, ranging from 5.48 to 7.01 dex; then, half of the galaxies in our sample have similar nitrogen abundances than the rest of the ELG (Figure 8(d)).

As with sulfur, argon is a product of rapid oxygen burning in the late stages of massive stars. It is a difficult element because two different ionization states can be detected in the galaxies' spectra in our sample, and only in eight galaxies, both lines ([Ar III]$\lambda$7136 and [Ar IV]$\lambda$4740) were detected. However, the abundances can be determined with only one of the lines (Izotov et al. 2006b; Pérez-Montero et al. 2007). Chemical abundances of argon were determined for 65 out of 85. When the $t_e$([Ar III]) was used to determine the abundance, the differences are about 0.15 dex, which are within 2$\sigma$. The argon abundances in our sample range from $12 + \log(Ar/H) = 4.89$ to 6.57, with an average value of 5.71 dex (Figure 8(e)). However, six of our objects present values lower than 5.3 dex; J021852.90-091218.7 is the galaxy with the lowest value, of 4.89 dex.

Finally, we can briefly discuss the abundances of chlorine and iron. There are only five and 12 galaxies where we could determine such abundances, respectively, and they are listed in columns (7) and (8) of Table 2. The average of the iron abundance is 5.62 dex, which is lower than for other galaxies (see Section 6.2). It could be claimed that our data were not corrected by depletion on grains. However, none of the other values were either. Rodríguez & Rubin (2005) obtained iron abundances for 11 galactic and extragalactic nebulae and late-type galaxies, and they cover from 5.70 up to 6.16 dex. The ICF cannot be blamed on this because we used the same that they did. One of the explanations for such deficiency of iron in these galaxies might be the lack of type Ia SN, which might be the most important source of iron. Finally, chlorine abundance could be determined in only five of the galaxies in our sample. Values range between 4.07 and 5.09 dex, with an average of 4.55 dex. They are similar to those obtained in some H II regions of late-type galaxies (Dufour 1984; Hidalgo-Gámez et al. 2001a).

Our results indicate that the EELGs, or oxygen-dominated galaxies, of this sample have subsolar abundances for all the elements.

### 5.1. log(X/O) versus Oxygen Abundance

In this section, we study the log(X/O) versus $12 + \log(O/H)$, where X is the chemical elements obtained from the spectra. As previously stated, the dispersion in the different log(X/O) plots were used to decide among the different ICFs, selecting those with the lowest dispersion. They are shown in Figure 9. Along with the data points, the median value of our sample (solid line) and the $\sigma$ (dotted lines) are shown. Chlorine, with only five data points, is shown just for completeness.

It is well known that the $\alpha$-elements (e.g., Ne, S, Ar), being primary elements, show a constant relationship with the oxygen abundance (Pagel 2009). However, in particular, sulfur and neon, show a trend for, at least, some range in metallicity. The former shows a decrease in values larger than 8.1 dex, except

Table 2
Chemical Abundances of the Different Elements Obtained in Our Sample of Oxygen-dominated Galaxies

| ID SDSS | (O/H) | (N/H) | (S/H) | (Ne/H) | (Ar/H) | (Cl/H) | (Fe/H) |
|---|---|---|---|---|---|---|---|
| J003218.60+150014.2 | 7.91 ± 0.13 | 5.91 ± 0.07 | 5.98 ± 0.04 | 7.32 ± 0.04 | 5.10 ± 0.02 | ⋯ | ⋯ |
| J004054.33+153409.7 | 8.09 ± 0.01 | 5.91 ± 0.06 | ⋯ | 7.65 ± 0.05 | ⋯ | ⋯ | ⋯ |
| J004236.93+160202.7 | 8.05 ± 0.04 | 6.69 ± 0.01 | ⋯ | 7.37 ± 0.07 | 6.08 ± 0.03 | ⋯ | ⋯ |
| J004529.15+133908.7 | 7.82 ± 0.03 | 6.85 ± 0.07 | ⋯ | 7.18 ± 0.04 | ⋯ | ⋯ | ⋯ |
| J005147.30+001940.0 | 7.74 ± 0.04 | 5.75 ± 0.06 | 5.85 ± 0.09 | 7.16 ± 0.03 | 5.32 ± 0.04 | ⋯ | 5.27 ± 0.18 |
| J010513.47−103741.0 | 8.20 ± 0.05 | 6.40 ± 0.03 | 6.29 ± 0.05 | 7.58 ± 0.08 | 5.82 ± 0.03 | ⋯ | ⋯ |
| J013344.63+005711.2 | 7.98 ± 0.01 | 5.80 ± 0.04 | ⋯ | 7.44 ± 0.02 | 5.53 ± 0.01 | ⋯ | ⋯ |
| J014721.68−091646.3 | 8.20 ± 0.06 | 6.33 ± 0.05 | ⋯ | 7.69 ± 0.14 | 5.86 ± 0.04 | ⋯ | ⋯ |
| J020051.59−084542.9 | 8.17 ± 0.05 | 6.46 ± 0.01 | ⋯ | 7.62 ± 0.07 | 5.86 ± 0.08 | ⋯ | ⋯ |
| J021852.90−091218.7 | 7.61 ± 0.02 | 5.82 ± 0.21 | 5.81 ± 0.31* | 6.94 ± 0.36 | 4.89 ± 0.04 | ⋯ | ⋯ |
| J022907.37−085726.2 | 8.20 ± 0.07 | 6.32 ± 0.01 | ⋯ | 7.70 ± 0.12 | 5.78 ± 0.07 | ⋯ | ⋯ |
| J024052.20−082827.4 | 7.72 ± 0.01 | 6.51 ± 0.03 | 5.98 ± 0.04 | 6.95 ± 0.04 | 5.55 ± 0.10 | 4.14 ± 0.08 | 5.43 ± 0.09 |
| J024453.67−082137.9 | 8.03 ± 0.02 | 6.12 ± 0.03 | 6.11 ± 0.03 | 7.40 ± 0.02 | 5.66 ± 0.01 | ⋯ | ⋯ |
| J025346.70−072344.1 | 7.86 ± 0.01 | 5.75 ± 0.03 | 6.14 ± 0.02* | 7.22 ± 0.06 | 5.66 ± 0.04 | ⋯ | ⋯ |
| J025436.11−000138.0 | 8.25 ± 0.06 | 6.40 ± 0.06 | ⋯ | ⋯ | 5.04 ± 0.03 | ⋯ | ⋯ |
| J030321.41−075923.2 | 7.69 ± 0.03 | 6.24 ± 0.06 | ⋯ | 6.92 ± 0.04 | 5.13 ± 0.09 | ⋯ | ⋯ |
| J030539.71−083905.3 | 8.06 ± 0.10 | 6.16 ± 0.02 | ⋯ | 7.52 ± 0.19 | 5.73 ± 0.05 | 4.86 ± 0.07 | ⋯ |
| J031623.96+000912.3 | 7.95 ± 0.02 | 6.32 ± 0.10 | 6.37 ± 0.02 | 7.31 ± 0.14 | 5.45 ± 0.08 | ⋯ | ⋯ |
| J032613.63−063513.5 | 8.02 ± 0.06 | 6.51 ± 0.01 | 6.36 ± 0.05 | 7.32 ± 0.09 | 5.83 ± 0.07 | ⋯ | ⋯ |
| J033031.22−005846.7 | 8.06 ± 0.03 | 6.30 ± 0.04 | 6.42 ± 0.15 | 7.41 ± 0.05 | 5.86 ± 0.02 | ⋯ | ⋯ |
| J033947.79−072541.3 | 8.12 ± 0.03 | 6.73 ± 0.03 | ⋯ | 7.55 ± 0.06 | 5.76 ± 0.05 | ⋯ | ⋯ |
| J040937.63−051805.8 | 8.01 ± 0.02 | 6.24 ± 0.04 | 6.19 ± 0.07 | 7.34 ± 0.08 | 5.75 ± 0.09* | ⋯ | ⋯ |
| J075315.74+401449.9 | 7.64 ± 0.04 | 5.84 ± 0.10 | ⋯ | 6.98 ± 0.08 | 5.31 ± 0.12 | ⋯ | ⋯ |
| J075715.74+452137.9 | 8.00 ± 0.07 | 6.45 ± 0.11 | ⋯ | 7.35 ± 0.08 | ⋯ | ⋯ | ⋯ |
| J080147.11+435302.3 | 8.24 ± 0.04 | 6.65 ± 0.06 | ⋯ | 7.65 ± 0.18 | 5.93 ± 0.06* | ⋯ | ⋯ |
| J082530.68+504804.5 | 8.09 ± 0.02 | 6.48 ± 0.01 | ⋯ | 7.47 ± 0.05 | 5.91 ± 0.02 | ⋯ | 5.58 ± 0.01 |
| J083350.24+454933.6 | 8.12 ± 0.05 | 6.80 ± 0.03 | 6.36 ± 0.05 | 7.49 ± 0.13 | 5.91 ± 0.04 | ⋯ | ⋯ |
| J083440.06+480540.9 | 8.05 ± 0.05 | 6.96 ± 0.09 | ⋯ | ⋯ | ⋯ | ⋯ | ⋯ |
| J084029.91+470710.2 | 7.46 ± 0.01 | 6.63 ± 0.05 | 5.68 ± 0.04 | 6.76 ± 0.09 | 5.28 ± 0.10 | ⋯ | 5.18 ± 0.18 |
| J084527.61+530853.0 | 8.16 ± 0.02 | 6.30 ± 0.07 | 6.23 ± 0.07 | 7.61 ± 0.09 | 5.86 ± 0.03 | ⋯ | 5.74 ± 0.17 |
| J085207.68−001118.0 | 8.27 ± 0.07 | 6.58 ± 0.02 | ⋯ | 7.75 ± 0.11 | 5.79 ± 0.04 | ⋯ | ⋯ |
| J090047.44+574255.2 | 8.25 ± 0.07 | 6.98 ± 0.01 | 6.23 ± 0.02 | 7.79 ± 0.17 | 6.09 ± 0.13 | ⋯ | ⋯ |
| J090122.81−002818.9 | 7.89 ± 0.02 | 6.02 ± 0.08 | ⋯ | 7.29 ± 0.09 | ⋯ | ⋯ | ⋯ |
| J090139.86+575946.2 | 8.22 ± 0.09 | 6.49 ± 0.02 | 6.23 ± 0.02 | 7.67 ± 0.15 | 5.96 ± 0.15 | ⋯ | ⋯ |
| J091652.24+003113.9 | 8.23 ± 0.05 | 6.85 ± 0.04 | 6.32 ± 0.16 | 7.67 ± 0.10 | 6.22 ± 0.07 | ⋯ | ⋯ |
| J092635.25+582047.3 | 8.12 ± 0.18 | 6.35 ± 0.29 | ⋯ | 7.61 ± 0.30 | 5.32 ± 0.23 | ⋯ | ⋯ |
| J092918.39+002813.2 | 7.94 ± 0.04 | 6.25 ± 0.06 | 6.16 ± 0.08 | 7.29 ± 0.05 | 5.76 ± 0.04 | 4.58 ± 0.06 | ⋯ |
| J093006.60+602653.0 | 7.91 ± 0.01 | 6.02 ± 0.04 | 6.05 ± 0.07* | 7.35 ± 0.12 | 5.65 ± 0.01 | ⋯ | 5.68 ± 0.10 |
| J094850.92+553716.1 | 7.91 ± 0.08 | 6.27 ± 0.05 | ⋯ | 7.23 ± 0.06 | ⋯ | ⋯ | ⋯ |
| J095023.32+004229.2 | 8.00 ± 0.02 | 6.59 ± 0.03 | 6.15 ± 0.07 | 7.40 ± 0.05 | 5.78 ± 0.03 | ⋯ | 5.84 ± 0.08 |
| J103344.05+635317.3 | 8.07 ± 0.05 | 6.85 ± 0.01 | ⋯ | 7.42 ± 0.06 | ⋯ | ⋯ | ⋯ |
| J104554.78+010405.8 | 8.09 ± 0.01 | 6.44 ± 0.01 | 6.22 ± 0.08 | 7.49 ± 0.14 | 5.90 ± 0.01 | 4.07 ± 0.20 | 5.99 ± 0.04 |
| J105032.49+661654.0 | 7.98 ± 0.02 | 5.99 ± 0.05 | 6.04 ± 0.14 | 7.40 ± 0.03 | 5.50 ± 0.01 | ⋯ | ⋯ |
| J112502.58−004525.6 | 8.01 ± 0.04 | 6.41 ± 0.11 | ⋯ | 7.41 ± 0.07 | ⋯ | ⋯ | ⋯ |
| J113303.80+651341.3 | 7.98 ± 0.05 | 6.28 ± 0.08 | ⋯ | 7.36 ± 0.06 | 5.72 ± 0.16 | ⋯ | ⋯ |
| J113341.19+634925.9 | 7.92 ± 0.02 | 5.92 ± 0.05 | 6.10 ± 0.06* | 7.34 ± 0.08 | 5.69 ± 0.02 | ⋯ | ⋯ |
| J113459.60−000104.2 | 8.04 ± 0.12 | 6.16 ± 0.26 | ⋯ | 7.42 ± 0.19 | ⋯ | ⋯ | ⋯ |
| J114649.34+005346.0 | 7.61 ± 0.01 | 5.73 ± 0.10 | ⋯ | 7.00 ± 0.03 | 5.31 ± 0.03 | ⋯ | ⋯ |
| J115247.52−004007.7 | 7.82 ± 0.02 | 5.93 ± 0.01 | 6.09 ± 0.17* | 7.15 ± 0.06 | 5.29 ± 0.10 | ⋯ | 5.14 ± 0.10 |
| J120055.64+032404.0 | 7.73 ± 0.02 | 6.07 ± 0.11 | ⋯ | 7.09 ± 0.08 | 5.41 ± 0.06 | ⋯ | ⋯ |
| J122419.73+010559.5 | 8.09 ± 0.04 | 6.10 ± 0.07 | 6.32 ± 0.11 | 7.55 ± 0.10 | 5.07 ± 0.01 | ⋯ | ⋯ |
| J123436.30−020721.3 | 8.30 ± 0.05 | 6.22 ± 0.08 | 6.21 ± 0.03 | 7.86 ± 0.11 | 5.82 ± 0.07 | ⋯ | ⋯ |
| J125526.06−021334.1 | 7.67 ± 0.01 | 5.88 ± 0.09 | 6.15 ± 0.09 | 6.98 ± 0.03 | 5.49 ± 0.05 | ⋯ | ⋯ |
| J130029.30+021502.9 | 7.56 ± 0.35 | 6.44 ± 0.10 | ⋯ | 6.89 ± 0.20 | ⋯ | ⋯ | ⋯ |
| J130148.03+013718.6 | 8.17 ± 0.08 | 6.54 ± 0.06 | 6.29 ± 0.06 | 7.61 ± 0.14 | 5.98 ± 0.12 | ⋯ | ⋯ |
| J130211.15−000516.4 | 8.19 ± 0.24 | 7.01 ± 0.18 | 6.81 ± 0.01 | 7.60 ± 0.18 | 6.32 ± 0.14 | ⋯ | ⋯ |
| J130249.19+653449.5 | 8.09 ± 0.03 | 6.05 ± 0.10 | 6.12 ± 0.06 | 7.61 ± 0.08 | 5.61 ± 0.02 | ⋯ | ⋯ |
| J131654.37−024930.4 | 7.66 ± 0.04 | 6.11 ± 0.14 | ⋯ | 6.83 ± 0.06 | ⋯ | ⋯ | 5.88 ± 0.01 |
| J131937.24+005043.8 | 8.34 ± 0.03 | 6.73 ± 0.06 | 6.31 ± 0.02 | 7.87 ± 0.15 | 6.15 ± 0.05 | ⋯ | ⋯ |
| J132347.46−013253.0 | 7.66 ± 0.01 | 5.48 ± 0.29 | 5.76 ± 0.02 | 6.93 ± 0.05 | 5.49 ± 0.02 | ⋯ | ⋯ |
| J132654.62+011346.8 | 7.97 ± 0.02 | 6.28 ± 0.08 | 6.08 ± 0.08 | 7.30 ± 0.03 | 5.71 ± 0.02 | ⋯ | ⋯ |
| J133303.96+624603.8 | 7.58 ± 0.11 | ⋯ | ⋯ | 7.11 ± 0.16 | ⋯ | ⋯ | ⋯ |
| J135155.90+032524.3 | 8.11 ± 0.07 | 6.97 ± 0.04 | 6.37 ± 0.06 | 7.52 ± 0.13 | 5.99 ± 0.10 | ⋯ | ⋯ |
| J135930.37−010322.2 | 8.33 ± 0.11 | 6.84 ± 0.07 | 6.32 ± 0.20 | 7.67 ± 0.25 | 6.57 ± 0.22 | ⋯ | ⋯ |

Table 2
(Continued)

| ID SDSS | (O/H) | (N/H) | (S/H) | (Ne/H) | (Ar/H) | (Cl/H) | (Fe/H) |
|---|---|---|---|---|---|---|---|
| J140725.31+052837.9 | 7.97 ± 0.06 | 6.17 ± 0.04 | 6.18 ± 0.01 | 7.34 ± 0.10 | 5.68 ± 0.06* | ⋯ | ⋯ |
| J141940.33+050906.8 | 8.13 ± 0.21 | ⋯ | ⋯ | 7.69 ± 0.33 | ⋯ | ⋯ | ⋯ |
| J142214.31−003919.5 | 7.93 ± 0.07 | 5.96 ± 0.08 | 6.30 ± 0.08 | 7.30 ± 0.10 | 5.50 ± 0.04 | ⋯ | ⋯ |
| J143804.21+013333.5 | 8.08 ± 0.07 | 6.90 ± 0.07 | ⋯ | 7.44 ± 0.18 | ⋯ | ⋯ | ⋯ |
| J144205.41−005248.6 | 8.13 ± 0.10 | 6.87 ± 0.07 | ⋯ | 7.49 ± 0.13 | 5.99 ± 0.01 | ⋯ | ⋯ |
| J144805.38−011057.7 | 7.95 ± 0.01 | 6.21 ± 0.03 | 6.24 ± 0.02 | 7.29 ± 0.02 | 5.82 ± 0.01 | ⋯ | 5.84 ± 0.04 |
| J152830.72+001740.2 | 8.11 ± 0.08 | 6.58 ± 0.08 | 6.59 ± 0.08 | 7.51 ± 0.16 | 6.01 ± 0.06 | ⋯ | ⋯ |
| J154108.38+032029.4 | 7.96 ± 0.03 | 6.25 ± 0.13 | ⋯ | 7.28 ± 0.06 | ⋯ | ⋯ | ⋯ |
| J154337.31−000608.1 | 8.01 ± 0.11 | ⋯ | ⋯ | ⋯ | ⋯ | ⋯ | ⋯ |
| J154654.55+030902.2 | 8.02 ± 0.01 | 6.33 ± 0.04 | ⋯ | 7.38 ± 0.04 | 5.69 ± 0.03 | ⋯ | ⋯ |
| J155644.29+540328.9 | 7.99 ± 0.09 | 6.19 ± 0.08 | 6.16 ± 0.04 | ⋯ | 5.83 ± 0.08 | 5.09 ± 0.09 | ⋯ |
| J164359.16+443633.0 | 8.22 ± 0.08 | 6.67 ± 0.03 | 6.16 ± 0.02 | 7.65 ± 0.15 | 6.01 ± 0.14 | ⋯ | ⋯ |
| J172009.83+542133.2 | 8.00 ± 0.06 | 5.84 ± 0.29 | ⋯ | 7.36 ± 0.14 | ⋯ | ⋯ | ⋯ |
| J172706.33+594902.2 | 8.02 ± 0.08 | 6.74 ± 0.04 | ⋯ | 7.31 ± 0.22 | ⋯ | ⋯ | ⋯ |
| J173126.55+591150.1 | 8.04 ± 0.08 | ⋯ | ⋯ | 7.36 ± 0.14 | ⋯ | ⋯ | ⋯ |
| J173501.25+570308.6 | 7.88 ± 0.01 | 6.63 ± 0.01 | 6.19 ± 0.10 | 7.15 ± 0.05 | 5.40 ± 0.02 | ⋯ | 5.88 ± 0.03 |
| J174327.60+544320.1 | 8.02 ± 0.11 | 6.27 ± 0.25 | 6.99 ± 0.07 | 7.35 ± 0.18 | ⋯ | ⋯ | ⋯ |
| J210114.40−055510.3 | 8.19 ± 0.02 | 6.47 ± 0.12 | ⋯ | 7.62 ± 0.10 | 5.86 ± 0.08 | ⋯ | ⋯ |
| J211527.07−075951.4 | 8.15 ± 0.01 | 6.64 ± 0.04 | 6.40 ± 0.01 | 7.55 ± 0.05 | 6.18 ± 0.03 | ⋯ | ⋯ |
| J232936.55−011056.9 | 8.08 ± 0.04 | 6.45 ± 0.02 | 6.31 ± 0.01 | 7.42 ± 0.09 | 5.76 ± 0.02 | ⋯ | ⋯ |
| J235347.69+005402.1 | 8.12 ± 0.10 | 6.34 ± 0.09 | ⋯ | 7.59 ± 0.15 | 5.78 ± 0.02 | ⋯ | ⋯ |

**Note.** (X/H) corresponds to canonical nomenclature as $12 + \log(X/H)$. Sulfur and argon abundances computed with our own $t_e$ from direct method are indicated with * (see Section 4.2.1).

(This table is available in machine-readable form.)

for two data points, which are the galaxies with the largest sulfur abundance. Such a negative trend is shown in other samples of ELG galaxies that use different ICFs (e.g., Kehrig et al. 2006; Hägele et al. 2008; Croxall et al. 2009; Izotov et al. 2011). Therefore, it cannot be an ICF artifact but likely due to depletion of O due to grains (Izotov et al. 2006).

The neon abundances also show an increment in the log(Ne/O) for oxygen abundances higher than 8.1 dex. Such trend was shown by Izotov et al. (2011) for their sample of luminous compact galaxies. Actually, their deviation is the same as our deviation from the median value. Another author, like Hägele et al. (2008), obtained a larger dispersion and an opposite trend.

Argon seems to show a large dispersion, increasing at high metallicities, around the median value. However, all except eight galaxies are in the $1\sigma$ locus, and, therefore, the log(Ar/O) is constant with the oxygen abundance. Other authors obtained a similar behavior, such as Croxall et al. (2009) and Hägele et al. (2008).

Therefore, none of the $\alpha$-elements, but argon, shows a constancy in the log(X/O) ratio with oxygen abundance. Pagel (2009) proposed two reasons for such lack of constancy: (i) an increment on the mass loss of high-metallicity massive stars, or (ii) "an increasing contribution from intermediate-stars at higher metallicity." However, our data cannot discriminate between these two statements.

Being iron, one of the elements that more easily forms dust grains, it is reasonable to find a clear trend of lower log(Fe/O) for higher oxygen abundances, as in other samples (e.g., Izotov et al. 2011). However, there is a constancy with a large dispersion around the median iron value, although the data points are very few for a conclusive answer.

Finally, nitrogen is always a very special element. There must be a trend with the abundance, at least for abundances larger than 8.2 dex (see, among others, Hidalgo-Gámez et al. 2012; and López-Sánchez & Esteban 2010b). However, we do not find such a trend here but a dispersion diagram, especially for lower abundances than 7.8 dex. Such a dispersion does not improve with any other ICF. As nitrogen is a secondary element, the ratio N/O might be influenced by the age of the star formation event, although there might be another parameter at work (Maiolino & Manucci 2019). If the star formation is quite recent, there has not been enough time for nitrogen to be released, and therefore, the N/O is small (e.g., Pilyugin 1993). However, for older star formation events, if nitrogen is released toward the interstellar medium, the N/O ratio increases.

## 6. Discussion

### 6.1. Comparison with Other Authors

We compared our results with previous works that obtained the abundances using an earlier data release (Kniazev et al. 2004). In Figure 10, we show the comparison of the oxygen abundances in the two studies.

Along with the data points, the 1:1 line is shown. Our abundances are lower than those ones in Kniazev et al. (2004) by 0.1 dex, on average, and our uncertainties are also smaller (see Section 2). Differences are larger at lower abundances, with only one of their measurements below $12 + \log(\text{O/H}) \sim 7.7$, which is the limit adopted to consider a galaxy as an XMP galaxy. Our measurements show that at least 10 galaxies can be considered XMPs due to their lower O/H.

Also, some other galaxies of our sample have been included in other investigations, and their oxygen abundances were previously determined. In total, 33 of the galaxies in our sample have oxygen abundances already determined, besides the values from Kniazev et al. (2004), and 11 of them have abundance of another element.

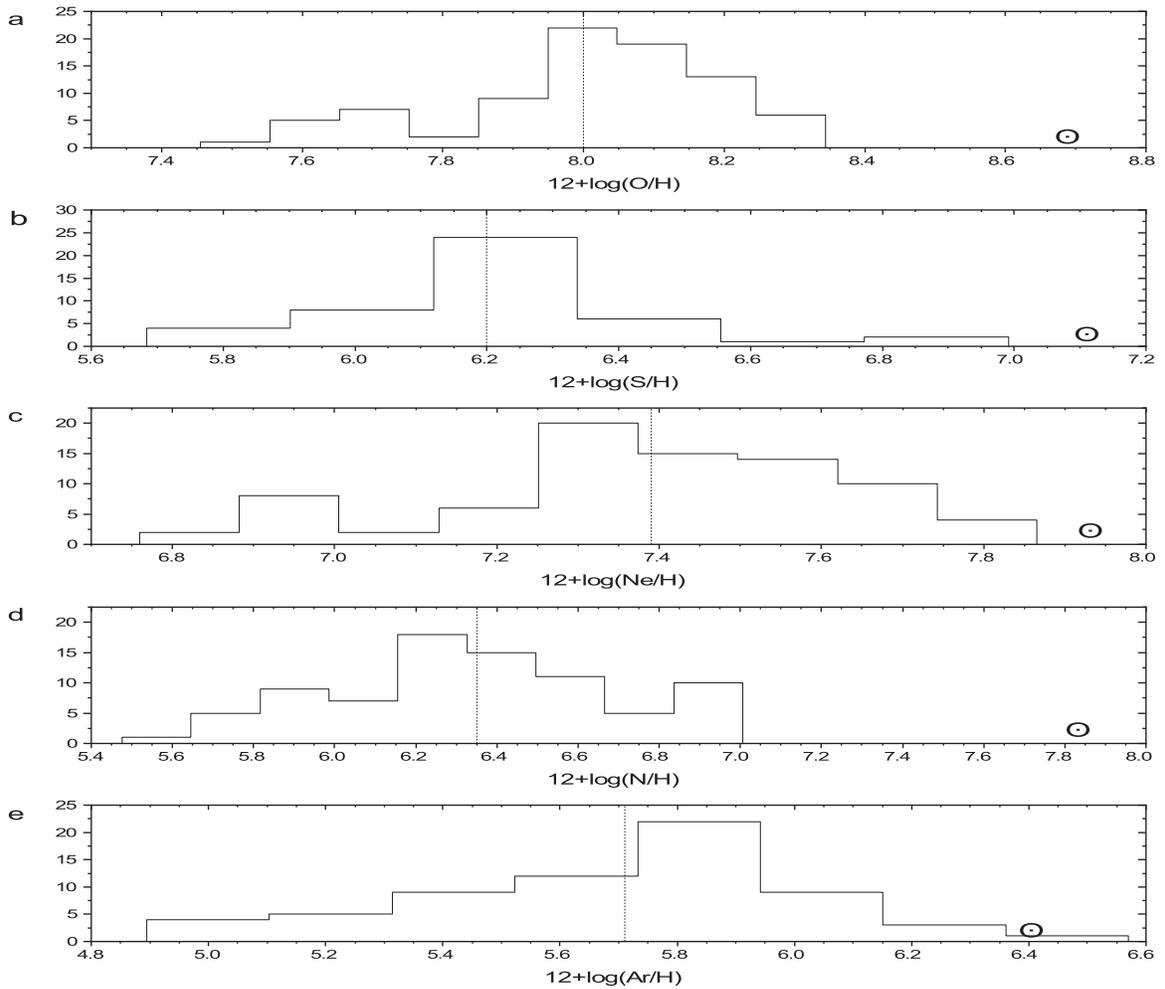

**Figure 8.** Distribution of the chemical abundances for the galaxies in our sample. The average value in each chemical abundance for the galaxies in the sample is indicated with the dashed line, and symbol ⊙ corresponds to solar value for each abundance (Asplund et al. 2009).

In Table 3, the chemical abundances of the 28 galaxies in common with other authors where the abundances were determined with the $T_e$ method are shown: 16 galaxies in common with Chávez et al. (2014; reference (1) in Table 3), five in common with Fernández et al. (2018; reference (2)), five in common with Izotov et al. (2011; reference (3)), three in common with Berg et al. (2019; reference (4)), two in common with Amorín et al. (2012; reference (5)), and another two in common with Loaiza-Agudelo et al. (2020; reference (6)). Finally, one galaxy is in common with Douglass & Vogeley (2017; reference (7)), and another is in common with Hägele et al. (2006; reference (8)). Of those, only two (2.27%) have abundances of all the elements studied here, and they are kept in our sample for completeness purposes. The common galaxies for C09 were omitted in Table 3 because the determination of oxygen abundances is with semiempirical methods, and their results are higher by 0.6 dex than those from the direct method, as noticed by Izotov et al. (2011). The galaxies in common with Ekta & Chengalur (2010), Mallery et al. (2007) were also not taken into account because they used SDSS data from earlier releases.

On average, our values are smaller for all the elements, but the differences are larger for nitrogen and sulfur. As shown, we used a different relation for the sulfur temperature determination (Equation (4)), which was particularly obtained for this kind of galaxies, resulting in larger $T_e$ values, and therefore lower abundances. Considering the nitrogen abundance, it has an important dependence on the ICF. The rest of our abundances are lower but still within the uncertainties.

### 6.2. Comparison with Other Emission-line Galaxy Samples

In this section, we compare the abundance values obtained here with the abundances obtained for ELGs and discuss possible differences in their chemical abundances. This can be summarized in Table 4, which shows the chemical abundance average for all the elements studied in different types of ELG. Along with our own data, the average values for dI include the SMC (Dufour 1984; Hidalgo-Gámez & Olofsson 2002), blue compact galaxies (BCGs; Hidalgo-Gámez & Olofsson 2002; Hägele et al. 2008), late spirals including LMC (Dufour 1984; Hidalgo-Gámez & Olofsson 2002), and some XMP (Papaderos et al. 2006). We also included the green peas galaxies (values from Hawley 2012 and Amorín et al. 2012), the KISS galaxies (Hirschauer et al. 2015), and DDO 68 (Annibali et al. 2019), which is the galaxy with the lowest metallicity known so far, and NGC 1705 (Annibali et al. 2015), which is another EELG according to the published spectrum.

According to the data in this table, all the ELGs (extreme or normal) have a very similar oxygen abundance, except XMP galaxies. Late spirals and KISS galaxies have a little more oxygen ($2\sigma$) than the rest of the galaxies, as expected. In any case, neither NGC 1705 nor the galaxies in our sample have lower oxygen abundance. Therefore, the large [O III] intensity is not a consequence of the low metallicity of the galaxy.

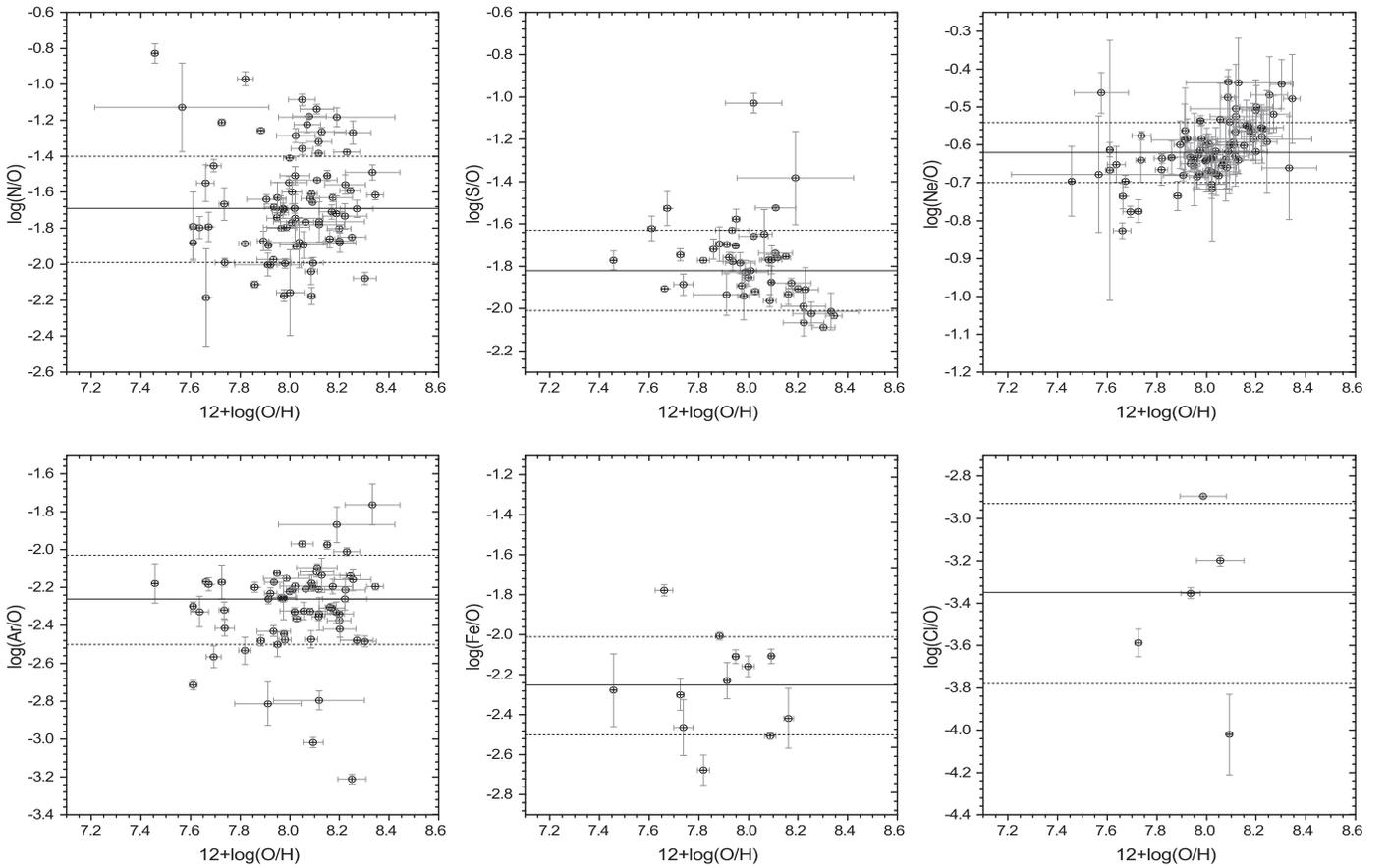

**Figure 9.** Abundance ratios vs. oxygen abundance $12 + \log(O/H)$ for our oxygen-dominated galaxies. The solid horizontal line is the median of our sample, and dashed lines are the standard deviation. All the distributions show a large dispersion.

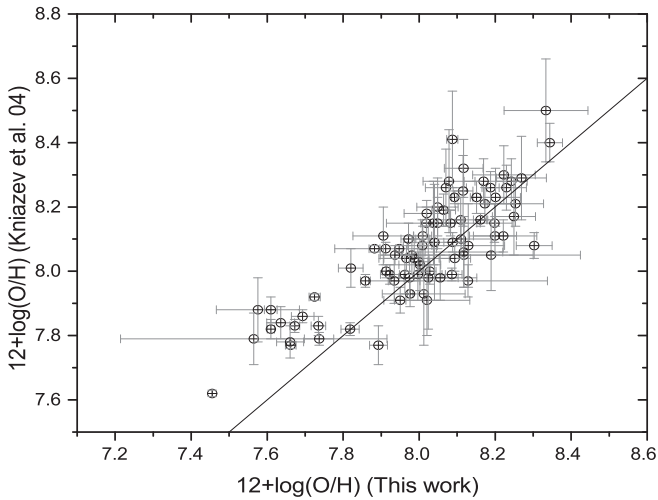

**Figure 10.** Comparison of $12 + \log(O/H)$ of the values presented in this work and those reported by Kniazev et al. (2004) for the oxygen-dominated galaxies. The solid line is the 1:1.

Neon and argon also have similar values for all the ELGs (except XMP galaxies), with a large dispersion. As these elements are released mainly from SNe II, it might be an indication that all of them have similar burst ages. This is expected because all the abundances were determined with the direct method, and in order to detect the forbidden oxygen line [O III]$\lambda$4363, the burst should be young.

The main differences are, again, the abundances of nitrogen, sulfur, and iron of the galaxies in our sample, which are the smallest of all the galaxies, except for the XMP. However, dI and BCG have very similar nitrogen abundances, and the differences in the sulfur abundances might be due to differences in the ICF and the $t_e$ used.

These low abundances might be related to the infall of gas. However, such infalls are very likely to be from type II SNe, so the enrichment might be on oxygen, neon, and argon, which have normal abundances. The infalls of primordial gas would dilute all the abundances (L. Carigi 2020, private communication). The enrichment winds from a previous generation of type Ia SNe would enhance the gas abundance of sulfur and iron, mainly (Pagel 2009).

The low abundances of iron can be explained by dust grains formation (Rodríguez & Rubin 2005), along with a low frequency of type Ia SNe in these galaxies because iron is profusely produced in these SNe (Carigi & Hernandez 2008; Franco & Carigi 2008). Also, sulfur is produced by these SNe mainly, with the time needed for the release of such elements being about 1 Gyr (Hernández-Martínez et al. 2011). Therefore, if the previous event of star formation was not long enough for such elements to be released, their abundances might be lower than those for other galaxies with more older star formation events. As a consequence, the galaxies with the larger $\log(Fe/O)$ and $\log(S/O)$ might be the ones with the older

**Table 3**
Differences in the Chemical Abundance of the EELGs of This Sample Compared to the Values Determined by Other Authors

| SDSS ID | 12 + log(O/H) This Work | 12 + log(O/H) Comp | 12 + log(S/H) This Work | 12 + log(S/H) Comp | 12 + log(Ne/H) This Work | 12 + log(Ne/H) Comp | 12 + log(N/H) This Work | 12 + log(N/H) Comp | 12 + log(Ar/H) This Work | 12 + log(Ar/H) Comp | Ref. |
|---|---|---|---|---|---|---|---|---|---|---|---|
| J003218.60 | 7.91 ± 0.13 | 8.02 ± 0.05 | 5.98 ± 0.04 | 6.61 ± 0.04 | ⋯ | ⋯ | 5.91 ± 0.07 | 6.55 ± 0.04 | 5.10 ± 0.02 | 5.66 ± 0.06 | 1, 2, 8 |
|  |  | 8.27 ± 0.07 |  | 6.49 ± 0.11 |  |  |  |  |  |  | 2, 8 |
|  |  | 7.93 ± 0.03 |  |  |  |  |  |  |  |  | 8 |
| J004054.33 | 8.09 ± 0.01 | 7.98 ± 0.06 | ⋯ | ⋯ | 7.65 ± 0.05 | 7.30 ± 0.03 | 5.91 ± 0.06 | 6.92 ± 0.02 | ⋯ | ⋯ | 5 |
|  |  | 8.29 ± 0.06 |  |  |  |  |  |  |  |  | 6 |
| J005147.30 | 7.74 ± 0.04 | 7.80 ± 0.05 | 5.85 ± 0.09 | 6.24 ± 0.5 | ⋯ | ⋯ | 5.75 ± 0.06 | 6.55 ± 0.05 | ⋯ | ⋯ | 1, 2 |
|  |  | 7.74 ± 0.04 |  |  |  |  |  |  |  |  | 2 |
| J013344.63 | 7.98 ± 0.01 | 8.03 ± 0.11 | ⋯ | ⋯ | ⋯ | ⋯ | ⋯ | ⋯ | ⋯ | ⋯ | 1 |
| J024052.20 | 7.72 ± 0.01 | 7.95 ± 0.13 | ⋯ | ⋯ | ⋯ | ⋯ | ⋯ | ⋯ | ⋯ | ⋯ | 1 |
| J024453.67 | 8.03 ± 0.02 | 7.99 ± 0.10 | ⋯ | ⋯ | ⋯ | ⋯ | ⋯ | ⋯ | ⋯ | ⋯ | 1 |
| J025346.70 | 7.86 ± 0.01 | 7.91 ± 0.02 | 6.14 ± 0.02 | 6.49 ± 0.04 | 7.22 ± 0.06 | 7.42 ± 0.05 | 5.75 ± 0.03 | 6.40 ± 0.04 | 5.66 ± 0.04 | 5.95 ± 0.04 | 4 |
| J030321.41 | 7.69 ± 0.03 | 7.89 ± 0.10 | ⋯ | ⋯ | ⋯ | ⋯ | ⋯ | ⋯ | ⋯ | ⋯ | 1 |
|  |  | 7.86 |  |  |  |  |  |  |  |  | 3 |
| J033947.79 | 8.12 ± 0.03 | 8.38 | ⋯ | ⋯ | ⋯ | ⋯ | ⋯ | ⋯ | ⋯ | ⋯ | 3 |
| J040937.63 | 8.01 ± 0.02 | 8.15 ± 0.10 | ⋯ | ⋯ | ⋯ | ⋯ | ⋯ | ⋯ | ⋯ | ⋯ | 1 |
| J082530.68 | 8.09 ± 0.02 | 8.10 ± 0.08 | ⋯ | ⋯ | ⋯ | ⋯ | ⋯ | ⋯ | ⋯ | ⋯ | 1 |
| J084029.91 | 7.46 ± 0.01 | 7.60 ± 0.02 | ⋯ | ⋯ | ⋯ | ⋯ | 6.63 ± 0.05 | 6.36 ± 0.08 | ⋯ | ⋯ | 2 |
| J091652.24 | 8.23 ± 0.05 | 8.32 ± 0.15 | ⋯ | ⋯ | ⋯ | ⋯ | ⋯ | ⋯ | ⋯ | ⋯ | 1 |
| J092918.39 | 7.94 ± 0.04 | 8.06 ± 0.17 | ⋯ | ⋯ | ⋯ | ⋯ | ⋯ | ⋯ | ⋯ | ⋯ | 1 |
| J093006.60 | 7.91 ± 0.01 | 8.02 ± 0.02 | 6.05 ± 0.07 | 6.41 ± 0.04 | 7.35 ± 0.12 | 7.59 ± 0.05 | 6.02 ± 0.04 | 6.55 ± 0.02 | 5.65 ± 0.01 | 5.84 ± 0.04 | 4 |
| J095023.32 | 8.00 ± 0.02 | 8.04 ± 0.13 | ⋯ | ⋯ | ⋯ | ⋯ | ⋯ | ⋯ | ⋯ | ⋯ | 1 |
| J104554.78 | 8.09 ± 0.01 | 8.17 ± 0.10 | ⋯ | ⋯ | ⋯ | ⋯ | ⋯ | ⋯ | ⋯ | ⋯ | 1 |
| J113303.80 | 7.98 ± 0.05 | 7.91 ± 0.10 | ⋯ | ⋯ | 7.36 ± 0.06 | 7.30 ± 0.06 | 6.28 ± 0.08 | 6.87 ± 0.03 | ⋯ | ⋯ | 5 |
|  |  | 7.97 |  |  |  |  |  |  |  |  | 3 |
| J130211.15 | 8.19 ± 0.24 | 8.12 | ⋯ | ⋯ | ⋯ | ⋯ | ⋯ | ⋯ | ⋯ | ⋯ | 3 |
| J130249.19 | 8.09 ± 0.03 | 8.05 ± 0.02 | ⋯ | ⋯ | ⋯ | ⋯ | 6.05 ± 0.10 | 6.58 ± 0.02 | ⋯ | ⋯ | 7, 8 |
| J132347.46 | 7.66 ± 0.01 | 7.58 ± 0.02 | ⋯ | ⋯ | 6.93 ± 0.05 | 7.06 ± 0.04 | 5.48 ± 0.29 | 6.25 ± 0.04 | 5.49 ± 0.02 | 5.50 ± 0.06 | 4 |
| J144805.38 | 7.95 ± 0.01 | 8.12 ± 0.12 | ⋯ | ⋯ | ⋯ | ⋯ | ⋯ | ⋯ | ⋯ | ⋯ | 1 |
| J172706.33 | 8.02 ± 0.08 | 8.24 | ⋯ | ⋯ | ⋯ | ⋯ | ⋯ | ⋯ | ⋯ | ⋯ | 3 |
| J173501.25 | 7.88 ± 0.01 | 8.11 ± 0.01 | 6.19 ± 0.10 | 6.54 ± 0.01 | ⋯ | ⋯ | 6.63 ± 0.01 | 7.28 ± 0.01 | ⋯ | ⋯ | 2 |
| J210114.40 | 8.19 ± 0.02 | 8.16 ± 0.20 | ⋯ | ⋯ | ⋯ | ⋯ | ⋯ | ⋯ | ⋯ | ⋯ | 1 |
| J211527.07 | 8.15 ± 0.01 | 8.23 ± 0.11 | 6.40 ± 0.01 | 6.53 ± 0.03 | ⋯ | ⋯ | 6.64 ± 0.04 | 7.12 ± 0.03 | ⋯ | ⋯ | 1, 2 |
|  |  | 8.25 ± 0.03 |  |  |  |  |  |  |  |  | 2 |
| J232936.55 | 8.08 ± 0.04 | 8.13 ± 0.16 | ⋯ | ⋯ | ⋯ | ⋯ | ⋯ | ⋯ | ⋯ | ⋯ | 1 |
| J235347.69 | 8.12 ± 0.10 | 8.19 ± 0.40 | ⋯ | ⋯ | ⋯ | ⋯ | ⋯ | ⋯ | ⋯ | ⋯ | 6 |

**Note.** See text for references.

(This table is available in machine-readable form.)

Table 4
Average Values of the Chemical Abundance of the EELGs in This Sample Compared to Average Values for Other Kind of Emission Line Late-type Galaxies

| Name | $12 + \log(O/H)$ | $12 + \log(S/H)$ | $12 + \log(Ne/H)$ | $12 + \log(N/H)$ | $12 + \log(Ar/H)$ | $12 + \log(Fe/H)$ | $12 + \log(Cl/H)$ | References |
|---|---|---|---|---|---|---|---|---|
| This Work | $8.00 \pm 0.06$ | $6.22 \pm 0.06$ | $7.39 \pm 0.10$ | $6.35 \pm 0.07$ | $5.71 \pm 0.06$ | $5.62 \pm 0.06$ | $4.67 \pm 0.08$ | |
| NGC 1705 (EELG) | $7.91 \pm 0.04$ | $6.34 \pm 0.05$ | $7.20 \pm 0.05$ | $6.63 \pm 0.03$ | $5.62 \pm 0.04$ | ... | ... | 1 |
| Green peas | $8.01 \pm 0.08$ | $6.66 \pm 0.45$ | $7.34 \pm 0.09$ | $6.87 \pm 0.08$ | $5.74 \pm 0.16$ | $6.33 \pm 0.30$ | ... | 2, 3 |
| dI (incl. SMC) | $7.93 \pm 0.09$ | $6.49 \pm 0.14$ | $7.25 \pm 0.85$ | $6.42 \pm 0.11$ | $5.78 \pm 0.12$ | ... | 4.7 | 4, 5 |
| BCGs | $7.98 \pm 0.17$ | $6.52 \pm 0.07$ | $7.28 \pm 0.22$ | $6.40 \pm 0.40$ | $5.75 \pm 0.06$ | $6.09 \pm 0.10$ | ... | 4, 6 |
| Late spiral (incl. LMC) | $8.06 \pm 0.30$ | $6.85 \pm 0.11$ | $7.47 \pm 0.30$ | $6.61 \pm 0.30$ | $6.20 \pm 0.06$ | ... | $4.84 \pm 0.20$ | 4, 5 |
| KISS galaxies | $8.10 \pm 0.04$ | $6.49 \pm 0.03$ | $7.43 \pm 0.05$ | $7.00 \pm 0.04$ | $5.64 \pm 0.05$ | ... | ... | 7 |
| XMP galaxies (this work) | $7.61 \pm 0.06$ | $5.85 \pm 0.12$ | $6.93 \pm 0.11$ | $6.02 \pm 0.13$ | $5.27 \pm 0.06$ | $5.53 \pm 0.09$ | ... | |
| XMP galaxies | $7.23 \pm 0.03$ | ... | $6.43 \pm 0.05$ | ... | ... | ... | ... | 8 |
| DDO 68 (XMP) | $6.96 \pm 0.09$ | $5.48 \pm 0.07$ | $6.31 \pm 0.05$ | $5.65 \pm 0.07$ | $4.76 \pm 0.07$ | ... | ... | 9 |

**References.** (1) Annibali et al. (2015); (2) Hawley (2012); (3) Amorín et al. (2012); (4) Hidalgo-Gámez & Olofsson (2002); (5) Dufour (1984); (6) Hägele et al. (2008); (7) Hirschauer et al. (2015); (8) Papaderos et al. (2006); (9) Annibali et al. (2019).

stellar population, because they had enough time to evolve for type Ia SNe to explode.

A similar situation might occur for nitrogen, which is produced mainly in stars with low masses, in particular in stars with masses between 4 and 8 $M_\odot$ (Chiappini et al. 2003). They need between 50 and 200 Myr to be released and cooled (Mollá et al. 2006). Then, if the age of the star formation events is smaller than 50 Myr, there is not enough time for nitrogen to be detected in the spectra, and the abundance is low.

Finally, there are not many measurements of the ratio of chlorine to oxygen in ELGs, but the dispersion of the data is also very broad (Hidalgo-Gámez et al. 2001a, 2001b). In any case, the values in late-type galaxies are smaller than those in galactic H II regions (Esteban et al. 2015).

From the results on the metallicity, we can conclude that the oxygen-dominant galaxies of this sample have not very different metallicity values from the rest of the ELGs included here. Only iron is clearly more deficient in our galaxies than in the rest of the late-type galaxies of our comparison sample that have iron abundance measurements, while nitrogen and sulfur are in the low content range. However, oxygen abundance is very similar to other late-type galaxies, and actually, the average is the same as the SMC oxygen abundance (8.02 dex, Dufour 1984). Therefore, the high intensity of the oxygen lines, especially the [O III]$\lambda$5007, is not the most determinant factor to decide whether a galaxy has a very low abundance.

### 6.3. Comparison with Models

We can compare our abundance results with chemical evolution models. We use those from Stasińska & Izotov (2003) because they used the simplest reasonable models. They are based on an axisymmetric, expanding cloud in the adiabatic phase with time variations of the density distribution, a Salpeter IMF with mass limits of 120 $M_\odot$ and 0.8 $M_\odot$, and an instantaneous burst of star formation. The data from our sample, divided in two metallicity bins, as in Stasińska & Izotov (2003), along with the curves for the two models that fit best the data are shown in Figure 11. Those best-fitting models are H3 for high-metallicity galaxies (black dots) and I5 for intermediate ones (red triangles). The I5 model includes X-ray photons as another source of ionization. However, none of these models fit perfectly all the ratios for both intermediate- and high-metallicity data. Although I5 model is the best fitting for our intermediate-metallicity data, it cannot fit the helium intensities (He II $\lambda$4686 and He I $\lambda$5875) or the [N II]/[O II] ratio (see Figure 11). However, it is the best approximation because the other models gave very low intensities, of the order of 0.01 or below while our intensities values are of the order of 0.05. Moreover, the existence of X-ray sources might explain the large values of EW([O III]) in, at least, part of the EELGs.

A more deep grasp of the large EW([O III]) line can be obtained from the effective temperature and the ionization parameter needed to get such values. It is well known that the larger the metallicity the lower the U (or the $T_{\rm ion}$) for the same intensity lines, at least for the simplest models (from the pioneering work from Campbell 1988, to Chisholm 2016). In Figure 12 (top), the intensity of the [O III]$\lambda$5007 versus Z is presented for a total of 642 galaxies. Along with the data points of our sample (diamonds), there are some other EELGs from the literature (crosses) and normal ELGs (small dots). The galaxies were selected from the references given in Table 4 along with data from Kniazev et al. (2004). Also, the lines represent the models from Campbell (1988): the solid line is a model with a $T_{\rm ion}$ of 38,200 K and an ionization parameter of 0.001, the dotted line is (65,600, 0.001) while the dashed line is (38,200, 0.1), and the dotted–dashed line is (65,600, 0.1). It is clear that in order to fit the intensity of [O III]$\lambda$5007 a high U, a high $T_{\rm ion}$, or both are needed. Although at high Z, a combination of low $T_{\rm ion}$ and high U can be considered for most of the EELGs, at low metallicity, only the high values of both parameters can fit the [O III]$\lambda$5007, as expected (e.g., Bresolin et al. 1999). On the contrary, most of the normal ELGs can be fitted with low values of both parameters. These results agree with the Stasińska & Izotov (2003) models and also with the explanation proposed by Steidel et al. (2014). In Figure 12 (bottom), the I([O III]$\lambda$4363) versus Z is shown. Although both lines come from the same ion, they do not show the same behavior. There is a clear correlation between the I([O III]$\lambda$4363) and the abundance with no differences between the two EELG samples while the relationship bends at about 7.7 dex for normal ELGs.

As said, in order to explain the larger intensities of the oxygen lines for EELGs, larger $T_{\rm ion}$ or low-ionization parameters are needed, although it is not for normal ELGs. Actually, the path of both kinds of galaxies differs, as can be seen in Figure 12. Such larger temperatures might be reached if other sources of ionization, besides photoionization, are considered as the models by Stasińska & Izotov (2003) suggest.

### 6.4. The Relation between Extreme Emission-line Galaxies and Extremely Metal-poor Galaxies

As previously said, some authors think that EELGs are just XMP galaxies. This is based on the idea that oxygen is one of the most important cooling elements in H II regions; therefore, the larger the intensity of the oxygen lines, the lower the abundance (e.g., Osterbrock & Ferland 2006). As an aside of this investigation, we will check carefully this statement to elucidate if this is the main reason for the large EW values.

From the results in the previous sections and Table 4, it is clear that the average abundances of all the elements of the galaxies in our sample are not similar to the abundances' values of the XMP galaxies. Actually, the oxygen abundance average of the EELGs in our sample is very similar to the SMC value or the dI and BCG value (Hidalgo-Gámez & Olofsson 2002), although most of the galaxies in Hidalgo-Gámez & Olofsson (2002) are not oxygen-dominated. Even for those elements with the lowest metallicity, as sulfur and nitrogen, the abundances for the galaxies in our sample differ a great deal from the abundances of XMP galaxies. Then, it can be concluded that not all the EELGs are of low abundances. Actually, only 10 of our galaxies have abundances smaller than 7.7 dex, and their average abundances, along with their dispersion, are shown in Table 4.

A quick comparison of the abundance of the XMP in this investigation with those XMP from Papaderos et al. (2006) or Annibali et al. (2019) shows that the average abundances of ours are higher than theirs (see Table 4). This might be due to the fact that only three galaxies in our sample have very low oxygen abundance, smaller than 7.6 dex. In any case, the XMP galaxies have the smallest abundances fobr all the chemical elements in our sample. We can go further and check this statement with the work by Filho et al. (2013), where they presented a sample of 141 XMP galaxies. However, only one-third are clearly oxygen-dominated galaxies. This ratio could be high, up to 50%, as in Sánchez Almeida et al. (2016). However, there is an important number of XMP galaxies that are not oxygen-dominated.

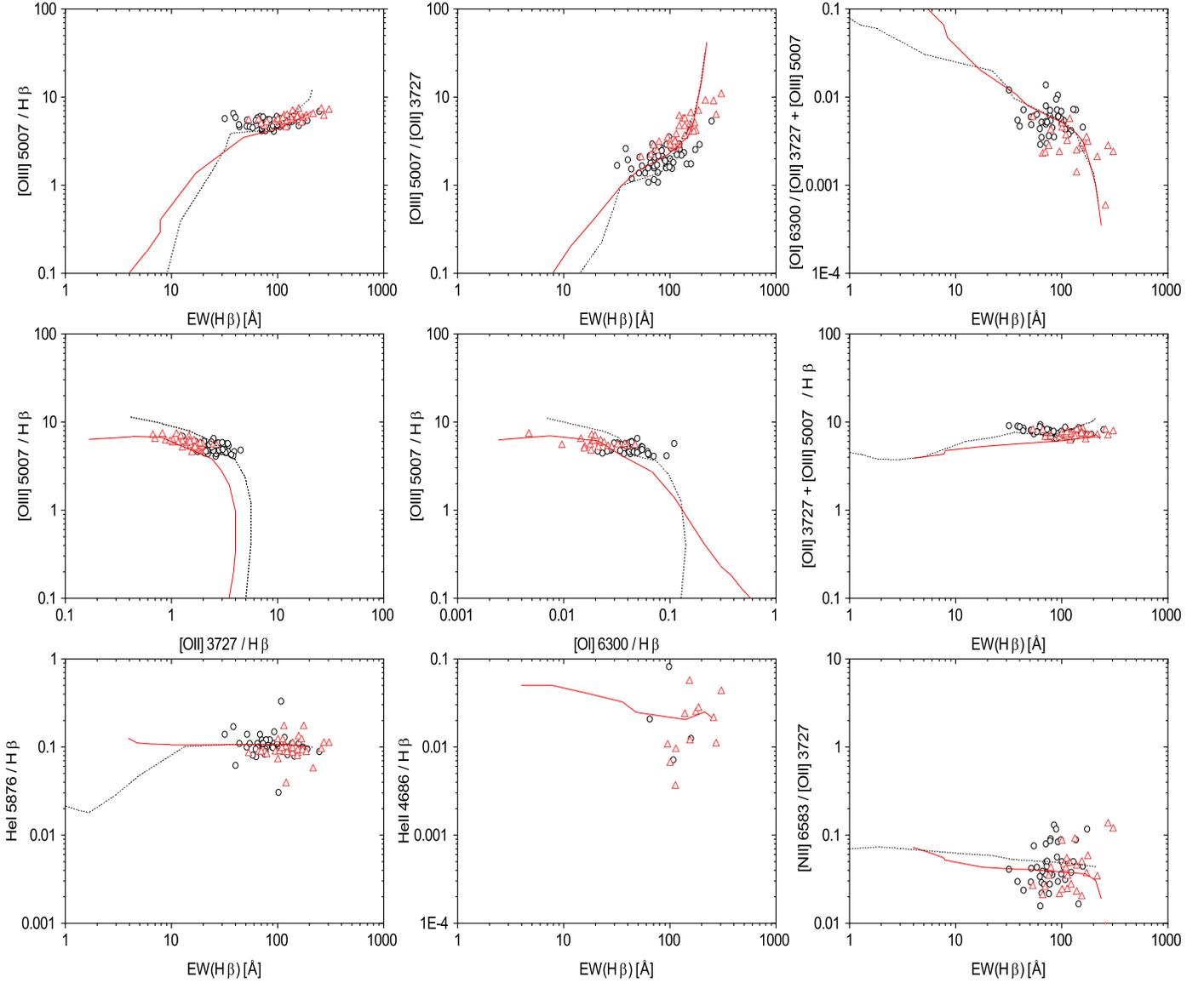

**Figure 11.** Emission-line diagrams for the galaxies from the sample. The black circles are the galaxies with high metallicity ($Z = 0.2 Z_\odot$), and red triangles are the ones with intermediate metallicity ($Z = 0.05 Z_\odot$). Overplotted are lines for the best-fitting curves that were taken from the evolutionary models from Stasińska & Izotov (2003). Those models were H3 (black line) and I5 (red line) for high and intermediate metallicity, respectively. All the I models included another ionization source (X-ray; see Stasińska & Izotov 2003 for details).

As stated, in our sample, only 12% of the EELGs are XMP. One reason might be the lack of completeness of sample (3). In order to check this, we have obtained several random subsamples (increasing the number of galaxies in each) from sample (2) and check how many XMP galaxies from sample (3) are in them. In all cases, the percentage of XMP galaxies was less than 17% with an average of 11%, indicating that, despite sample (3) not containing all the EELGs at low-z, the number of XMP galaxies does not depend on the completeness of the sample, with similar percentages (approximately 10%) of XMP from whatever the EELG sample is. This is in agreement with the values obtained for other EELG galaxies at different redshifts (Amorín et al. 2014; Ly et al. 2014, 2016; Calabrò et al. 2017). In Amorin et al. (2015), only 4% of their galaxies are XMP, which is even a smaller percentage than ours.

Finally, if the low metallicity is the reason for high EW and intensities values, a strong correlation is expected between these two parameters. Such relationship is shown in Figure 13. The correlation is not that strong, as expected ($r_g = 0.4$) with a broad dispersion. It is clear that low-metallicity galaxies (<7.8 dex) have larger EW (the average EW([O III]) for low-metallicity galaxies is 1169 Å, while it is 616 Å for those that are large metallicity). However, there are enough galaxies with EW([O III]) larger than 500 Å, but with no low abundances.

In summary, it can be concluded that the majority of the EELGs at low and intermediate redshift are not XMP, as can be seen in Table 2, although there is a tendency for XMP to be EELGs. We can now investigate those EELGs at high redshift. The first thing to check is if both samples are similar. According to Maseda et al. (2014), the spectral characteristic, which are the only ones we can compare to, of their galaxies are [O III]/H$\beta$ > 5 and [NII]/H $\alpha \ll 1$. Then, the galaxies presented here are spectroscopically similar to these and, therefore, comparable. Although the number of XMP galaxies at the high-$z$ sample is larger (33% compared to 10% at low-$z$), it is still too small to say that low chemical abundances are the reason why these galaxies have such large [O III]$\lambda$5007 intensities and EW.

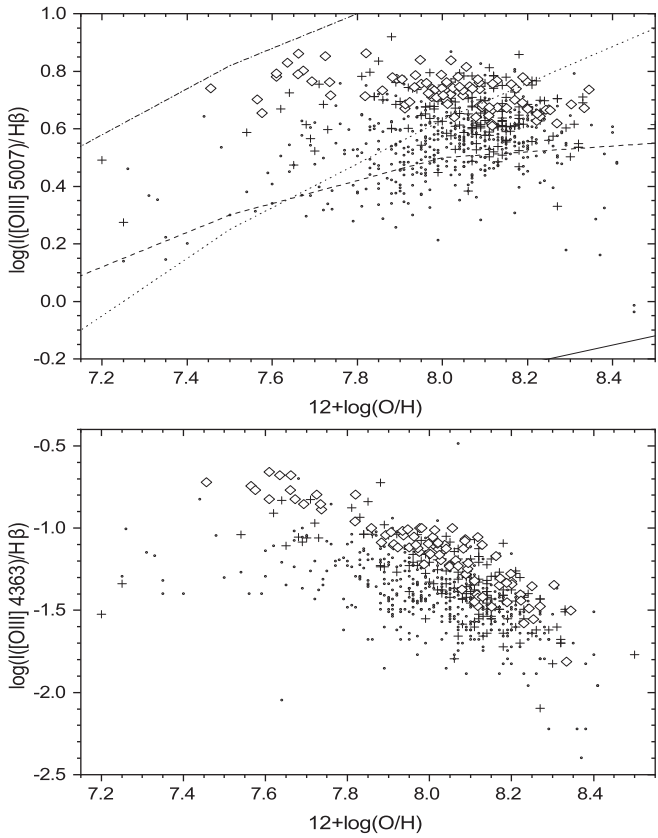

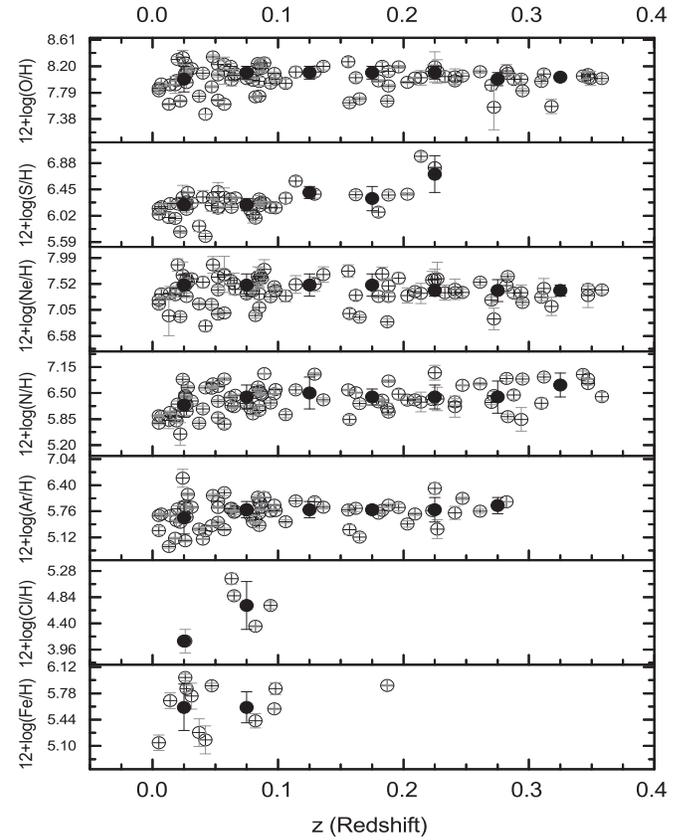

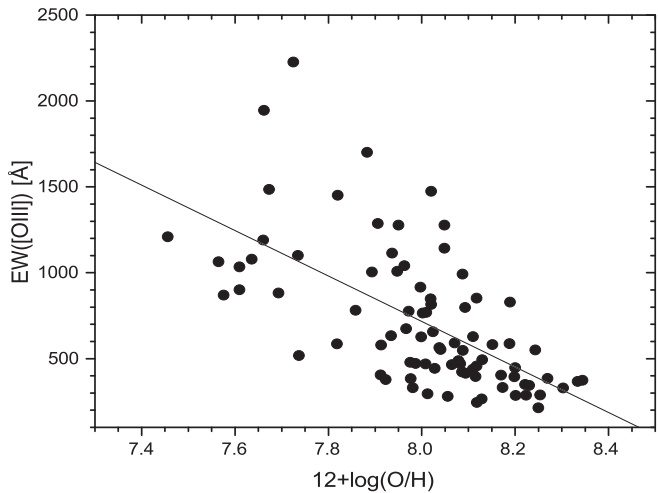

**Figure 12.** Oxygen-line ratios [O III]$\lambda$5007/H$\beta$ (top panel) and [O III]$\lambda$4363/H$\beta$ (bottom panel) vs. oxygen abundance. Data points of our sample (diamonds), EELGs (crosses), and normal ELGs (small dots). Lines represent the models from Campbell (1988; see text).

**Figure 13.** Relationship between the EW([O III]) and $12 + \log(O/H)$ for the oxygen-dominated galaxies in our sample. It is clear that the dispersion grows for lower abundances.

**Figure 14.** Element abundances and redshift of the galaxies from the sample. There is no trend on these parameters, showing there is no evolution process in this kind of galaxies. Filled circles represent the average abundance by a redshift bin of 0.5.

### 6.5. Chemical Evolution with Time

It is expected that galaxies at high redshift might have lower metallicities. This might be due to a pollution of the gas with metals because of the increases of the SFR events. Although our sample is at low redshift, some insight can be obtained from the variation of the chemical abundances with the redshift. This is shown in Figure 14 for all the elements, although both chloride and iron are not detected in galaxies farther than $z \approx 0.1$, and they are only shown for the sake of completeness. Along with the data points, the average abundance of the galaxies, without the XMP, for redshift bins of 0.5 are shown as solid dots. The first thing to notice in this figure is the large dispersion, with very different abundances for the same redshift. The second interesting fact is that the abundance of all the elements is very constant with small decrements at very low redshift, lower than 0.05. Sulphur shows an increases in its abundance at high $z$, but this value is determined from only two data points; therefore, it is not significant.

It is clear that there is a lack of evolution in the chemical abundance for all the elements for such a long time, about 4.6 Gyr. Although it is possible they suffer a strong evolution in earlier times ($0.35 < z < 2$) that polluted all the gas, it is not very likely that it will occur for every single galaxy in such a large sample as presented here. Another possibility is a continuous SFR, which will increase the metal content of the gas slowly or several burst of star formation at different times, as Amorín et al. (2012) proposed for green peas. Although the best way to check the chemical evolution is modeling the star formation history, some clues can be guessed from the spectra. It is well known that when a galaxy has an underlying old stellar population the continuum spectra is more prominent (e.g., Cairós et al. 2001; Gil de Paz et al. 2003). However, all the galaxies in sample (3) show a very weak stellar continuum. This might indicate that the number of star formation episodes are not that many, and, then, the chemical evolution cannot be that strong as needed for the large enrichment required in order to increase the metallicity a factor of 10, from 10% to 50% solar abundance. Neither of our selection criteria (the detection of the [O III]$\lambda$4363 or the composite color) might have any influence on the lack of evolution because the [O III]$\lambda$4363 line is related to the age of the present burst of star formation but not with the history of the star formation while the *gri* color is related with the redshift only but not with the stellar population of the galaxy.

**Table 5**
Average Values of the Physical Properties and Chemical Abundances of the Different Elements in Oxygen-dominated Galaxies through Their Different Colors

| Color | EW (Å) | $n_e$(cm$^{-3}$) | $t_e$([O III]) | $t_e$([O II]) | (O/H) | (N/H) | (S/H) | (Ne/H) | (Ar/H) | (Fe/H) |
|---|---|---|---|---|---|---|---|---|---|---|
| Blue | 754 | 159 ± 25 | 1.38 ± 0.57 | 1.37 ± 0.05 | 7.98 ± 0.03 | 6.17 ± 0.07 | 6.11 ± 0.08 | 7.36 ± 0.10 | 5.59 ± 0.04 | 5.59 ± 0.11 |
| Purple | 566 | 161 ± 12 | 1.27 ± 0.13 | 1.29 ± 0.11 | 8.05 ± 0.05 | 6.38 ± 0.05 | 6.23 ± 0.06 | 7.45 ± 0.10 | 5.79 ± 0.06 | 5.62 ± 0.06 |
| Green | 812 | 350 ± 293 | 1.36 ± 0.18 | 1.34 ± 0.12 | 7.98 ± 0.07 | 6.43 ± 0.10 | 6.50 ± 0.05 | 7.35 ± 0.11 | 5.73 ± 0.08 | 5.88 ± 0.01 |

Because of this dependence, we can use the different *gri* colors to check the previous results, although this selection is highly artificial (Izotov et al. 2011). Also, there is such extensive literature, on *blueberry galaxies* and especially *green peas*, that we decided to study the different colors in detail. The average values of the physical properties and chemical abundances of these subsamples are shown in Table 5.

There are slight differences between the three colored subsamples in the average values: namely the density, the EW, and the abundances of nitrogen, sulfur, argon, and iron. Normally, the blue galaxies have the lowest abundances (N, S, and Ar), while the green ones have the highest ones. Actually, if only the green galaxies are considered in the comparison with the green peas of Table 4, the differences in N and S are much smaller. Though, there is not a significant tendency of one group to be deficient (or overabundant) in all the values. There is also differences in the distribution of the abundances, mainly for argon and neon (see Hidalgo-Gámez & Miranda-Pérez 2017).

It must be reminded that the more distant galaxies are those from the green subsample with $z > 0.15$; therefore, some of the spectra from these green galaxies showed low S/N, and it is possible to overestimate (underestimate) the measure of the intensities. This might be important for the density; because, in spectra with lower S/N, it is more difficult to separate the doublet at [S II]$\lambda\lambda$6717, 6731 from where $n_e$ is obtained. Also, the [S III]$\lambda$9069 line is absent in these galaxies because the redshift pushes this line outside the wavelength range observed. So, perhaps, the lack of this line overestimates the sulfur abundance.

From this table, it can be concluded that the *gri* composite color is not a defining characteristic of the galaxies, and they should not be selected by them, only.

So, the lack of chemical evolution seems to be real although it is not possible to know such evolution from the data presented in this investigation. Of course, a more detail study is needed to get a more conclusive answer, but such study is beyond the scope of this investigation.

## 7. Summary and Conclusions

We have studied a sample of oxygen-dominated galaxies, which are those ELGs where the double-ionized oxygen line at 5007 Å is the most prominent line of the optical spectrum, instead of H$\alpha$. They are also called EELGs. We selected only those galaxies that have the oxygen auroral line at 4363 Å, detected with enough S/N for a reliable determination of the chemical abundances. Such restriction might exclude galaxies farther than $z = 0.35$. As we used integrated spectra, compact galaxies were preferred, between 2″ and 4″ in Petrossian radius. Finally, we included only those galaxies with unusual colors, such as green, purple, and bright blue. In total, we studied 88 EELGs.

We determined the electron density from the sulfur doublet as well as the electron temperature from [O III] for all the galaxies in the sample. Although most of the galaxies have normal densities, typical in normal ELGs, there are some with values larger than 200 cm$^{-3}$. Also the electron temperature from [S III] could be determined for five galaxies and for three galaxies from [Ar III]. Very few galaxies have large temperature values ($T$[O III] larger than 15,000 K). The [S III] temperature is larger than the [O III] one, so when the sulfur lines were not present, the [S III] temperatures were determined with a model.

The chemical abundances of oxygen, neon, and nitrogen were determined for almost all the galaxies in the sample, while the abundances of sulfur and argon were determined for at least two-thirds of the sample. The iron and chlorine abundances were determined for 12 and five galaxies, respectively. None of the physical or chemical properties, except the density and sulfur, iron, and nitrogen abundance, differ much from the values in other late-type galaxies. Actually, the average values of the chemical abundances are very similar to those of the SMC for most of the elements. One reason for the lower N, S, and Fe abundances might be a young burst of star formation in these EELGs, in comparison with older ones in other ELGs.

It is very interesting to notice that only 10 out of 85 galaxies in the sample have oxygen abundances smaller than 7.70 dex; hence, it cannot be considered that the low abundances are the reason for the large EW([O III]). Regarding the XMP galaxies subsample, the average values of chemical abundances of all elements, including iron, are low compared to the total sample. However, they are not as low as those of very low-metallicity galaxies, (e.g., DDO 68; Annibali et al. 2019). In conclusion, we can say that oxygen-dominated galaxies with very large EW([O III]) have chemical abundances very similar to other late-type galaxies, such as dIrr or BCDs.

The best explanation for the large EW and intensities of the [O III]$\lambda$5007 line is the one proposed by Steidel et al. (2014) where the main reason for the large EW([O III]) is a hard spectrum in addition to a high-ionization parameter and, possibly, a high density. We notice that the galaxies in our sample have high-ionization lines, high densities and high-ionization parameters along with X-ray sources, all of them proposed by these authors as the typical characteristics of a highly ionized galaxy.

Finally, we noticed that the chemical abundance is very constant with the redshift, at least for intermediate values. Therefore, we can conclude that there has not been much chemical evolution of these galaxies in the last $5 \times 10^9$ yr.

## Acknowledgments

We would like to thank the anonymous referees for valuable comments and suggestions that have improved the manuscript. This manuscript is partly based on the Ph.D. Thesis of B.E. Miranda-Pérez, financed by CONACyT. The authors wish to thank Mrs J. Arellano and Mr. P. Gómez-Garcés for a careful revision of the English grammar, and to thank Á.R. López-Sánchez, L. Carigi, and H. Castañeda (who always will be remembered) for helpful discussions. Also thanks to Á.R. López-Sánchez for a careful reading of the manuscript. This manuscript has been financed by the Instituto Politécnico Nacional under research projects SIP-20220744 and SIP-20230583 projects. The manuscript is based on observations made with Sloan Digital Sky Survey (https://www.sdss.org), at Apache Point Observatory in New Mexico, United States.

*Facilities:* SDSS, ESO.

*Software:* ESO-MIDAS (European Southern Observatory 2013).

## Appendix

The content of Tables 6–13 is available in its entirety in the electronic edition of the *Astrophysical Journal*.

Table 6
Corrected Relative Line Intensities from 3705.04 to 3868.75 Å

| ID SDSS | 3721.83(Å) [S III] | 3721.94(Å) H14 | 3727.00(Å) [O II] | 3750.15(Å) H12 | 3770.63(Å) H11 | 3784.89(Å) He I | 3797.90(Å) H10 | 3805.78(Å) He I | 3819.61(Å) He I | 3825.00(Å) | 3835.39(Å) H9 | 3856.02(Å) Si II | 3868.75(Å) [Ne III] |
|---|---|---|---|---|---|---|---|---|---|---|---|---|---|
| J003218.60 | ⋯ | ⋯ | ⋯ | ⋯ | 2.37 ± 0.40 | ⋯ | 2.64 ± 0.52 | ⋯ | ⋯ | ⋯ | 5.55 ± 0.51 | ⋯ | 54 ± 1 |
| J004054.33 | ⋯ | ⋯ | 313 ± 3 | ⋯ | ⋯ | ⋯ | ⋯ | ⋯ | ⋯ | ⋯ | 12 ± 1 | ⋯ | 90 ± 1 |
| J004236.93 | ⋯ | ⋯ | 204 ± 2 | 2.13 ± 0.04 | 3.08 ± 0.05 | ⋯ | 6.89 ± 0.09 | ⋯ | ⋯ | ⋯ | 6.02 ± 0.08 | ⋯ | 45 ± 1 |
| J004529.15 | ⋯ | ⋯ | 112 ± 1 | ⋯ | ⋯ | ⋯ | ⋯ | ⋯ | 5.97 ± 0.10 | ⋯ | 6.21 ± 0.10 | ⋯ | 80 ± 1 |
| J005147.30 | ⋯ | ⋯ | 152 ± 2 | 2.49 ± 0.71 | 2.85 ± 0.69 | ⋯ | 3.81 ± 0.94 | ⋯ | ⋯ | ⋯ | 6.73 ± 0.78 | ⋯ | 66 ± 1 |
| J010513.47 | ⋯ | ⋯ | 270 ± 3 | ⋯ | ⋯ | ⋯ | ⋯ | ⋯ | ⋯ | ⋯ | 3.43 ± 0.07 | ⋯ | 56 ± 1 |
| J013344.63 | ⋯ | ⋯ | 250 ± 1 | ⋯ | ⋯ | ⋯ | ⋯ | ⋯ | 4.81 ± 0.51 | ⋯ | ⋯ | ⋯ | 72 ± 1 |
| J014721.68 | ⋯ | ⋯ | 335 ± 3 | 0.71 ± 0.02 | 4.20 ± 0.06 | ⋯ | 4.96 ± 0.06 | ⋯ | ⋯ | ⋯ | 5.04 ± 0.06 | 2.56 ± 0.04 | 59 ± 1 |
| J020051.59 | ⋯ | ⋯ | 268 ± 3 | ⋯ | 0.52 ± 0.02 | ⋯ | 1.89 ± 0.04 | ⋯ | ⋯ | ⋯ | ⋯ | ⋯ | 56 ± 1 |
| J021852.90 | ⋯ | ⋯ | ⋯ | ⋯ | ⋯ | ⋯ | ⋯ | ⋯ | ⋯ | ⋯ | 8.69 ± 1.56 | ⋯ | 69 ± 3 |

**Note.** Tables 6–13 are available in their entirety in the electronic edition of the *Astrophysical Journal*. Only the first ten galaxies are shown here for guidance regarding its form and content.

(This table is available in its entirety in machine-readable form.)

**Table 7**
Corrected Relative Line Intensities from 3871.82 to 4416.27 Å

| ID SDSS | 3871.82(Å) He I | 3888.00(Å) He I+H8 | 3926.53(Å) He I | 3964.73(Å) He I | 3967.00(Å) [Ne III]+H7 | 4075.86(Å) O II | 4101.74(Å) H$\delta$ | 4153.30(Å) O II | 4267.15(Å) C II | 4303.82(Å) O II | 4340.47(Å) H$\gamma$ | 4363.21(Å) [O III] | 4368.25(Å) O I |
|---|---|---|---|---|---|---|---|---|---|---|---|---|---|
| J003218.60 | ⋯ | 25 ± 1 | ⋯ | 1.38 ± 0.42 | 31 ± 1 | ⋯ | 29 ± 1 | ⋯ | ⋯ | ⋯ | 53 ± 1 | 7.85 ± 0.42 | ⋯ |
| J004054.33 | ⋯ | 26 ± 1 | ⋯ | 3.72 ± 0.04 | 46 ± 1 | ⋯ | 34 ± 1 | ⋯ | ⋯ | ⋯ | 47 ± 1 | 8.54 ± 0.09 | ⋯ |
| J004236.93 | ⋯ | 24 ± 1 | ⋯ | ⋯ | 32 ± 1 | ⋯ | 29 ± 1 | ⋯ | ⋯ | ⋯ | 52 ± 1 | 5.94 ± 0.08 | ⋯ |
| J004529.15 | ⋯ | 22 ± 1 | ⋯ | ⋯ | 44 ± 1 | ⋯ | 32 ± 1 | ⋯ | ⋯ | 7.88 ± 0.11 | 56 ± 1 | 16 ± 1 | ⋯ |
| J005147.30 | ⋯ | 26 ± 1 | ⋯ | ⋯ | 37 ± 1 | ⋯ | 30 ± 1 | ⋯ | ⋯ | ⋯ | 53 ± 1 | 13 ± 1 | ⋯ |
| J010513.47 | ⋯ | 19 ± 1 | ⋯ | ⋯ | 30 ± 1 | ⋯ | 24 ± 1 | ⋯ | ⋯ | ⋯ | 48 ± 1 | 5.27 ± 0.07 | ⋯ |
| J013344.63 | ⋯ | 22 ± 1 | ⋯ | ⋯ | 36 ± 1 | ⋯ | 28 ± 1 | ⋯ | ⋯ | ⋯ | 53 ± 1 | 10 ± 1 | ⋯ |
| J014721.68 | 4.49 ± 0.06 | 22 ± 1 | ⋯ | ⋯ | 33 ± 1 | ⋯ | 30 ± 1 | ⋯ | ⋯ | ⋯ | 51 ± 1 | 4.71 ± 0.06 | ⋯ |
| J020051.59 | ⋯ | 16 ± 1 | ⋯ | ⋯ | 25 ± 1 | ⋯ | 27 ± 1 | ⋯ | ⋯ | ⋯ | 48 ± 1 | 4.48 ± 0.06 | ⋯ |
| J021852.90 | ⋯ | 20 ± 2 | ⋯ | ⋯ | 38 ± 3 | ⋯ | 28 ± 2 | ⋯ | ⋯ | ⋯ | 48 ± 3 | 15 ± 2 | ⋯ |

(This table is available in its entirety in machine-readable form.)

**Table 8**
Corrected Relative Line Intensities from 4471.09 to 4985.90 Å

| ID SDSS | 4471.09(Å) He I | 4576.00(Å) O II | 4658.10(Å) N II | 4676.24(Å) O II | 4686.24(Å) He II | 4701.53(Å) [Fe II] | 4711.37(Å) [Ar IV] | 4740.16(Å) [Ar IV] | 4861.33(Å) H$\beta$ | 4905.34(Å) [Fe II] | 4921.93(Å) He II | 4958.91(Å) [O III] | 4985.90(Å) Fe III |
|---|---|---|---|---|---|---|---|---|---|---|---|---|---|
| J003218.60 | 3.28 ± 0.43 | ⋯ | ⋯ | ⋯ | ⋯ | ⋯ | ⋯ | ⋯ | 100 ± 1 | ⋯ | ⋯ | 161 ± 1 | 1.62 ± 0.21 |
| J004054.33 | 11 ± 1 | ⋯ | ⋯ | ⋯ | ⋯ | ⋯ | ⋯ | ⋯ | 100 ± 1 | 7.44 ± 0.07 | ⋯ | 187 ± 2 | ⋯ |
| J004236.93 | ⋯ | ⋯ | ⋯ | ⋯ | ⋯ | ⋯ | ⋯ | ⋯ | 100 ± 1 | ⋯ | ⋯ | 175 ± 2 | ⋯ |
| J004529.15 | ⋯ | ⋯ | ⋯ | ⋯ | ⋯ | ⋯ | ⋯ | ⋯ | 100 ± 1 | ⋯ | 1.75 ± 0.05 | 267 ± 3 | ⋯ |
| J005147.30 | 3.62 ± 0.34 | ⋯ | 0.61 ± 0.33 | ⋯ | 1.06 ± 0.29 | ⋯ | 1.96 ± 0.33 | 0.77 ± 0.25 | 100 ± 1 | ⋯ | 1.79 ± 0.31 | 176 ± 1 | 1.02 ± 0.21 |
| J010513.47 | ⋯ | ⋯ | ⋯ | ⋯ | ⋯ | ⋯ | ⋯ | ⋯ | 100 ± 1 | ⋯ | ⋯ | 196 ± 2 | ⋯ |
| J013344.63 | 4.53 ± 0.14 | ⋯ | ⋯ | ⋯ | ⋯ | ⋯ | ⋯ | ⋯ | 100 ± 1 | ⋯ | ⋯ | 180 ± 1 | ⋯ |
| J014721.68 | 4.37 ± 0.05 | ⋯ | ⋯ | ⋯ | ⋯ | ⋯ | ⋯ | ⋯ | 100 ± 1 | ⋯ | 7.66 ± 0.09 | 160 ± 2 | ⋯ |
| J020051.59 | 2.87 ± 0.04 | ⋯ | ⋯ | ⋯ | ⋯ | ⋯ | ⋯ | ⋯ | 100 ± 1 | ⋯ | ⋯ | 169 ± 2 | ⋯ |
| J021852.90 | 4.22 ± 1.34 | ⋯ | ⋯ | ⋯ | 1.18 ± 0.70 | ⋯ | 2.18 ± 0.73 | ⋯ | 100 ± 4 | ⋯ | ⋯ | 205 ± 6 | ⋯ |

(This table is available in its entirety in machine-readable form.)

**Table 9**
Corrected Relative Line Intensities from 4996.98 to 5512.77 Å

| ID SDSS | 4996.98(Å) O II | 5006.84(Å) [O III] | 5015.68(Å) He I | 5041.03(Å) Si II | 5047.74(Å) He I | 5084.77(Å) [Fe III] | 5121.83(Å) C II | 5191.82(Å) [Ar III] | 5200.26(Å) [N I] | 5261.61(Å) [Fe II] | 5355.65(Å) [Fe II] | 5412.00(Å) [Fe III] | 5512.77(Å) O I |
|---|---|---|---|---|---|---|---|---|---|---|---|---|---|
| J003218.60 | ⋯ | 468 ± 1 | 1.69 ± 0.18 | ⋯ | ⋯ | ⋯ | ⋯ | ⋯ | ⋯ | ⋯ | ⋯ | ⋯ | 0.89 ± 0.20 |
| J004054.33 | ⋯ | 546 ± 6 | ⋯ | ⋯ | ⋯ | ⋯ | ⋯ | ⋯ | ⋯ | ⋯ | ⋯ | ⋯ | ⋯ |
| J004236.93 | ⋯ | 521 ± 5 | ⋯ | ⋯ | ⋯ | ⋯ | ⋯ | ⋯ | ⋯ | ⋯ | ⋯ | ⋯ | ⋯ |
| J004529.15 | ⋯ | 729 ± 7 | 4.09 ± 0.06 | ⋯ | ⋯ | ⋯ | ⋯ | ⋯ | ⋯ | ⋯ | ⋯ | ⋯ | ⋯ |
| J005147.30 | ⋯ | 520 ± 1 | ⋯ | 1.62 ± 0.31 | ⋯ | ⋯ | ⋯ | ⋯ | ⋯ | ⋯ | ⋯ | ⋯ | ⋯ |
| J010513.47 | ⋯ | 581 ± 6 | ⋯ | ⋯ | ⋯ | ⋯ | ⋯ | ⋯ | ⋯ | ⋯ | ⋯ | ⋯ | ⋯ |
| J013344.63 | ⋯ | 549 ± 1 | 1.85 ± 0.12 | ⋯ | ⋯ | ⋯ | ⋯ | ⋯ | ⋯ | ⋯ | ⋯ | ⋯ | ⋯ |
| J014721.68 | ⋯ | 470 ± 5 | ⋯ | ⋯ | ⋯ | ⋯ | ⋯ | ⋯ | ⋯ | ⋯ | ⋯ | ⋯ | ⋯ |
| J020051.59 | ⋯ | 502 ± 5 | ⋯ | ⋯ | ⋯ | ⋯ | ⋯ | ⋯ | ⋯ | ⋯ | ⋯ | ⋯ | ⋯ |
| J021852.90 | ⋯ | 604 ± 3 | ⋯ | ⋯ | ⋯ | ⋯ | ⋯ | ⋯ | ⋯ | ⋯ | ⋯ | ⋯ | ⋯ |

(This table is available in its entirety in machine-readable form.)

**Table 10**
Corrected Relative Line Intensities from 5537.88 to 6462.00 Å

| ID SDSS | 5537.88(Å) [Cl III] | 5698.21(Å) [C III] | 5710.76(Å) N II | 5875.64(Å) He I | 5907.21(Å) C II | 5941.68(Å) N II | 6000.20(Å) [Ni III] | 6046.44(Å) O I | 6300.30(Å) [O I] | 6312.10(Å) [S II] | 6363.78(Å) [O I] | 6454.80(Å) C II | 6462.00(Å) C II |
|---|---|---|---|---|---|---|---|---|---|---|---|---|---|
| J003218.60 | ⋯ | ⋯ | ⋯ | 8.76 ± 0.31 | ⋯ | ⋯ | ⋯ | ⋯ | 1.82 ± 0.14 | 0.83 ± 0.08 | ⋯ | ⋯ | ⋯ |
| J004054.33 | ⋯ | 0.92 ± 0.01 | ⋯ | 7.84 ± 0.08 | ⋯ | ⋯ | ⋯ | ⋯ | ⋯ | ⋯ | ⋯ | ⋯ | ⋯ |
| J004236.93 | ⋯ | ⋯ | ⋯ 10 ± 1 | ⋯ | ⋯ | ⋯ | ⋯ | ⋯ | ⋯ | ⋯ | ⋯ | ⋯ | ⋯ |
| J004529.15 | ⋯ | ⋯ | ⋯ | 13 ± 1 | ⋯ | ⋯ | ⋯ | ⋯ | ⋯ | ⋯ | ⋯ | ⋯ | 12 ± 1 |
| J005147.30 | ⋯ | ⋯ | ⋯ | 8.73 ± 0.22 | ⋯ | ⋯ | ⋯ | ⋯ | 1.58 ± 0.16 | 0.94 ± 0.16 | ⋯ | ⋯ | ⋯ |
| J010513.47 | ⋯ | ⋯ | ⋯ | 11 ± 1 | ⋯ | ⋯ | ⋯ | ⋯ | 4.42 ± 0.05 | 1.25 ± 0.02 | ⋯ | ⋯ | ⋯ |
| J013344.63 | ⋯ | ⋯ | ⋯ | 8.71 ± 0.16 | ⋯ | ⋯ | ⋯ | ⋯ | 1.79 ± 0.07 | ⋯ | ⋯ | ⋯ | ⋯ |
| J014721.68 | ⋯ | ⋯ | ⋯ | 9.47 ± 0.11 | ⋯ | ⋯ | ⋯ | ⋯ | 4.97 ± 0.07 | ⋯ | ⋯ | ⋯ | ⋯ |
| J020051.59 | ⋯ | ⋯ | ⋯ | 9.62 ± 0.11 | ⋯ | ⋯ | ⋯ | ⋯ | 3.80 ± 0.05 | ⋯ | ⋯ | ⋯ | ⋯ |
| J021852.90 | ⋯ | ⋯ | ⋯ | 8.39 ± 0.68 | ⋯ | ⋯ | ⋯ | ⋯ | 1.86 ± 0.48 | 1.25 ± 0.36 | ⋯ | ⋯ | ⋯ |

(This table is available in its entirety in machine-readable form.)

**Table 11**
Corrected Relative Line Intensities from 6527.10 to 7115.40 Å

| ID SDSS | 6527.10(Å) [N II] | 6548.03(Å) [N II] | 6562.21(Å) Hα | 6583.41(Å) [N II] | 6678.15(Å) He I | 6716.47(Å) [S II] | 6730.85(Å) [S II] | 6769.61(Å) N I | 6791.25(Å) Ne II | 6855.88(Å) He I | 7002.23(Å) O I | 7083.00(Å) Ar I | 7115.40(Å) Si I |
|---|---|---|---|---|---|---|---|---|---|---|---|---|---|
| J003218.60 | ⋯ | 1.18 ± 0.11 | 286 ± 1 | 6.53 ± 0.19 | 2.02 ± 0.13 | 11 ± 1 | 8.00 ± 0.17 | ⋯ | ⋯ | ⋯ | ⋯ | ⋯ | ⋯ |
| J004054.33 | ⋯ | ⋯ | 286 ± 3 | 5.21 ± 0.05 | ⋯ | 4.70 ± 0.05 | 4.83 ± 0.05 | ⋯ | ⋯ | ⋯ | ⋯ | ⋯ | ⋯ |
| J004236.93 | ⋯ | 5.76 ± 0.07 | 286 ± 2 | 24 ± 1 | ⋯ | 13 ± 1 | 8.64 ± 0.09 | ⋯ | ⋯ | ⋯ | ⋯ | ⋯ | ⋯ |
| J004529.15 | ⋯ | ⋯ | 286 ± 3 | ⋯ | 2.06 ± 0.05 | 2.11 ± 0.04 | 2.64 ± 0.05 | ⋯ | ⋯ | ⋯ | ⋯ | ⋯ | ⋯ |
| J005147.30 | ⋯ | 1.12 ± 0.17 | 286 ± 1 | 3.19 ± 0.20 | 2.21 ± 0.16 | 6.01 ± 0.17 | 4.62 ± 0.20 | ⋯ | ⋯ | ⋯ | ⋯ | 1.72 ± 0.17 | ⋯ |
| J010513.47 | ⋯ | 2.39 ± 0.03 | 286 ± 3 | 7.80 ± 0.09 | 3.47 ± 0.04 | 15 ± 1 | 11 ± 1 | ⋯ | ⋯ | ⋯ | ⋯ | 2.12 ± 0.03 | ⋯ |
| J013344.63 | ⋯ | 1.03 ± 0.06 | 286 ± 1 | 5.13 ± 0.08 | 1.69 ± 0.06 | 9.80 ± 0.09 | 6.99 ± 0.09 | ⋯ | ⋯ | ⋯ | ⋯ | 0.72 ± 0.09 | ⋯ |
| J014721.68 | ⋯ | 3.32 ± 0.04 | 286 ± 2 | 9.37 ± 0.11 | 2.04 ± 0.03 | 17 ± 1 | 13 ± 1 | 1.97 ± 0.03 | 8.42 ± 0.10 | ⋯ | ⋯ | ⋯ | ⋯ |
| J020051.59 | ⋯ | 2.27 ± 0.03 | 286 ± 2 | 10 ± 1 | 2.02 ± 0.03 | 13 ± 1 | 9.54 ± 0.11 | ⋯ | ⋯ | ⋯ | ⋯ | 1.27 ± 0.02 | ⋯ |
| J021852.90 | ⋯ | 0.94 ± 0.36 | 286 ± 2 | 2.46 ± 0.47 | 2.13 ± 0.41 | 5.57 ± 0.48 | 4.09 ± 0.47 | ⋯ | ⋯ | ⋯ | ⋯ | 1.73 ± 0.42 | ⋯ |

(This table is available in its entirety in machine-readable form.)

**Table 12**
Corrected Relative Line Intensities from 7135.78 to 7751.10 Å

| ID SDSS | 7135.78(Å) [Ar III] | 7155.14(Å) [Fe II] | 7160.58(Å) He I | 7231.12(Å) C II | 7298.05(Å) He I | 7318.39(Å) [O II] | 7330.73(Å) [O II] | 7377.83(Å) [NiII] | 7411.61(Å) [NiII] | 7423.64(Å) N I | 7452.54(Å) [Fe II] | 7530.54(Å) [Cl IV] | 7751.10(Å) [Ar III] |
|---|---|---|---|---|---|---|---|---|---|---|---|---|---|
| J003218.60 | 4.90 ± 0.19 | ⋯ | ⋯ | ⋯ | ⋯ | ⋯ | ⋯ | ⋯ | ⋯ | ⋯ | ⋯ | ⋯ | ⋯ |
| J004054.33 | ⋯ | ⋯ | ⋯ | ⋯ | ⋯ | ⋯ | ⋯ | ⋯ | ⋯ | ⋯ | ⋯ | ⋯ | ⋯ |
| J004236.93 | 9.25 ± 0.10 | ⋯ | ⋯ | ⋯ | ⋯ | ⋯ | ⋯ | ⋯ | ⋯ | ⋯ | 5.54 ± 0.06 | ⋯ | ⋯ |
| J004529.15 | ⋯ | ⋯ | ⋯ | ⋯ | ⋯ | ⋯ | ⋯ | ⋯ | ⋯ | ⋯ | ⋯ | ⋯ | ⋯ |
| J005147.30 | 3.02 ± 0.18 | ⋯ | ⋯ | ⋯ | 0.50 ± 0.18 | ⋯ | ⋯ | ⋯ | ⋯ | ⋯ | ⋯ | ⋯ | ⋯ |
| J010513.47 | 4.07 ± 0.06 | ⋯ | ⋯ | ⋯ | ⋯ | ⋯ | 2.34 ± 0.03 | ⋯ | ⋯ | ⋯ | ⋯ | ⋯ | ⋯ |
| J013344.63 | 4.41 ± 0.09 | ⋯ | ⋯ | ⋯ | ⋯ | ⋯ | ⋯ | ⋯ | ⋯ | ⋯ | ⋯ | ⋯ | ⋯ |
| J014721.68 | 5.10 ± 0.07 | ⋯ | ⋯ | ⋯ | ⋯ | ⋯ | ⋯ | 1.41 ± 0.03 | ⋯ | 1.27 ± 0.02 | 8.95 ± 0.11 | ⋯ | ⋯ |
| J020051.59 | 4.51 ± 0.06 | ⋯ | ⋯ | ⋯ | ⋯ | ⋯ | ⋯ | ⋯ | ⋯ | ⋯ | ⋯ | ⋯ | ⋯ |
| J021852.90 | 4.01 ± 0.47 | ⋯ | ⋯ | ⋯ | ⋯ | ⋯ | ⋯ | ⋯ | ⋯ | ⋯ | ⋯ | ⋯ | ⋯ |

(This table is available in its entirety in machine-readable form.)

Table 13
Corrected Relative Line Intensities from 7816.13 to 9069.90 Å

| ID SDSS | 7816.13(Å) He I | 8030.69(Å) Ar II | 8094.08(Å) He I | 8125.30(Å) [CrII] | 8271.93(Å) H I | 8703.25(Å) N I | 8776.77(Å) He I | 8829.40(Å) [S III] | 8899.00(Å) He I | 9069.90(Å) [S III] | c(H$\beta$) |
|---|---|---|---|---|---|---|---|---|---|---|---|
| J003218.60 | ⋯ | ⋯ | ⋯ | ⋯ | ⋯ | ⋯ | ⋯ | ⋯ | ⋯ | ⋯ | 0.181 ± 0.002 |
| J004054.33 | ⋯ | ⋯ | ⋯ | ⋯ | ⋯ | ⋯ | ⋯ | ⋯ | ⋯ | ⋯ | 0.704 ± 0.001 |
| J004236.93 | ⋯ | ⋯ | ⋯ | ⋯ | ⋯ | ⋯ | ⋯ | ⋯ | ⋯ | ⋯ | 0.459 ± 0.001 |
| J004529.15 | ⋯ | ⋯ | ⋯ | ⋯ | ⋯ | ⋯ | ⋯ | ⋯ | ⋯ | ⋯ | 0.475 ± 0.001 |
| J005147.30 | ⋯ | ⋯ | ⋯ | ⋯ | ⋯ | ⋯ | 0.53 ± 0.19 | ⋯ | ⋯ | ⋯ | 0.090 ± 0.001 |
| J010513.47 | ⋯ | ⋯ | ⋯ | ⋯ | ⋯ | ⋯ | ⋯ | ⋯ | ⋯ | ⋯ | 0.012 ± 0.002 |
| J013344.63 | ⋯ | ⋯ | ⋯ | ⋯ | ⋯ | ⋯ | ⋯ | ⋯ | ⋯ | ⋯ | 0.198 ± 0.001 |
| J014721.68 | ⋯ | ⋯ | ⋯ | ⋯ | ⋯ | ⋯ | ⋯ | ⋯ | ⋯ | ⋯ | 0.143 ± 0.001 |
| J020051.59 | ⋯ | ⋯ | ⋯ | ⋯ | ⋯ | ⋯ | ⋯ | ⋯ | ⋯ | ⋯ | 0.253 ± 0.001 |
| J021852.90 | ⋯ | ⋯ | ⋯ | ⋯ | ⋯ | 0.60 ± 0.39 | ⋯ | 1.41 ± 0.45 | ⋯ | 5.99 ± 0.37 | 0.338 ± 0.003 |

(This table is available in its entirety in machine-readable form.)

## ORCID iDs

B. E. Miranda-Pérez  https://orcid.org/0000-0003-2409-3195

A. M. Hidalgo-Gámez  https://orcid.org/0000-0002-6596-2070